\documentclass[reprint,article,superscriptaddress,msmath,amssymb,aps]{revtex4-1}
\usepackage{graphicx}
\usepackage{dcolumn}
\usepackage{bm}
\usepackage{color}
\usepackage[]{natbib}
\usepackage{amsmath,amssymb,amsfonts,mathtools}
\usepackage{float}
\usepackage{pdfcomment}
\usepackage{verbatim}
\usepackage[utf8]{inputenc}
\usepackage{booktabs,chemformula}

\begin{document}
\preprint{APS/123-QED}
\title{Room temperature ferroic orders in Zr and (Zr, Ni) doped SrTiO\textsubscript{3} }

\author{Shahran Ahmed}
\affiliation{Department of Electrical and Electronic Engineering, University of Dhaka, Dhaka-1000, Bangladesh}
\author{A. K. M. Sarwar Hossain Faysal}
\affiliation{Department of Electrical and Electronic Engineering, University of Dhaka, Dhaka-1000, Bangladesh}
\author{M. N. I. Khan}\thanks{\texttt{ni\_khan}77@yahoo.com}
\affiliation{Materials Science Division, Atomic Energy Centre, Dhaka-1000, Bangladesh}
\author{M. A. Basith} \thanks{mabasith@phy.buet.ac.bd}
\affiliation{Nanotechnology Research Laboratory, Department of Physics, Bangladesh University of Engineering and Technology, Dhaka-1000, Bangladesh}
\author{Muhammad Shahriar Bashar}
\affiliation{Institute of Fuel Research and Development, Bangladesh Council of Scientific and Industrial Research, Dhaka-1205, Bangladesh }
\author{H. N. Das}
\affiliation{Materials Science Division, Atomic Energy Centre, Dhaka-1000, Bangladesh}
\author{Tarique Hasan}
\affiliation{Department of Electrical and Electronic Engineering, University of Dhaka, Dhaka-1000, Bangladesh}

\author{Imtiaz Ahmed}\thanks{imtiaz@du.ac.bd}
\affiliation{Department of Electrical and Electronic Engineering, University of Dhaka, Dhaka-1000, Bangladesh}

\begin{abstract}
We synthesized strontium titanate SrTiO\textsubscript{3} (STO), Zr doped $\text{Sr}_\text{1-x}\text{Zr}_\text{x}\text{Ti}\text{O}_3$ and (Zr, Ni) co-doped $\text{Sr}_\text{1-x}\text{Zr}_\text{x}\text{Ti}_\text{1-y}\text{Ni}_\text{y}\text{O}_3$ samples using solid state reaction technique to report their structural, electrical and magnetic properties. The cubic $Pm$-$3m$ phase of the synthesized samples has been confirmed using Rietveld analysis of the powder X-ray diffraction pattern. The grain size of the synthesized materials was reduced significantly due to Zr doping as well as (Zr, Ni) co-doping in STO. The chemical species of the samples were identified using energy-dispersive X-ray spectroscopy. We observed forbidden first order Raman scattering at 148, 547 and 797 cm$^{-1}$ which may indicate nominal loss of inversion symmetry in cubic STO. The absence of absorption at 500 cm$^{-1}$ and within 600-700 cm$^{-1}$ band in Fourier Transform Infrared spectra corroborates Zr and Ni as substitutional dopants in our samples. Due to 4\% Zr doping in $\text{Sr}_\text{0.96}\text{Zr}_\text{0.04}\text{Ti}\text{O}_3$ sample dielectric constant, remnant electric polarization, remnant magnetization and coercivity were increased. Notably, in the case of 4\% Zr and 10\% Ni co-doping we have observed clearly the existence of both FE and FM hysteresis loops in Sr\textsubscript{0.96}Zr\textsubscript{0.04}Ti\textsubscript{0.90}Ni\textsubscript{0.10}O\textsubscript{3} sample. In this co-doped sample, the remnant magnetization and coercivity were increased by $\sim$1 and $\sim$2 orders of magnitude respectively as compared to those of undoped STO. The coexistence of FE and FM orders in (Zr, Ni) co-doped STO might have the potential for interesting multiferroic applications.

 

\end{abstract}

\maketitle

\section{Introduction}
Strontium titanate  SrTiO\textsubscript{3} (STO) is perhaps one of the most prominent prototypical ABO\textsubscript{3} perovskite oxide materials having potentials in diverse fields of applications such as high-k dielectrics \cite{robertson2000band}, substrate materials \cite{christensen2019stimulating}, tunable microwave devices \cite{eriksson2003orientation, fuchs1999high}, non-volatile memory applications \cite{szot2006switching}, O\textsubscript{2} and H\textsubscript{2}S sensing \cite{hu2004new, kajale2012synthesis} and photocatalytic water splitting \cite{sakata2016photocatalytic}, to name just a few. At room temperature, STO exists in cubic structure with space group $Pm$-$3m$ where the $\text{Sr}^{2+}$ occupies A site which is surrounded by four TiO\textsubscript{6} octahedra and $\text{Ti}^{4+}$ occupies B site which is located at the octahedral void formed by six $\text{O}^{2-}$ ions situated at the faces of the cube \cite{li2004formability}. Pure stoichiometric STO is a band-insulator whose quantum paraelectric behaviour excludes the emergence of the ferroic phases such as ferroelectric (FE) and ferromagnetic (FM) orders \cite{spaldin2019advances, kustov2020domain, muller1979srti, haeni2004room, choudhury2011tuning}. Over the years numerous attempts have been made to induce FE and FM orders in STO by means of cation doping for achieving multiferroic behaviour. The FM ordering is induced in STO by doping transition metals (Mn, Fe, Co, Cd) and rare-earth elements (Yb, Mn)\cite{yao2011raman, ahmed2019enhancement, padmini2019investigation, muralidharan2014carrier, norton2002properties, lee2003magnetic}. The FE behavior has been observed at room temperature in Pr doped STO \cite{duran2005ferroelectricity} whereas Mn doping introduced FE relaxor type behaviour at low temperature \cite{tkach2005structure, choudhury2011tuning, choudhury2013magnetization, azzoni2000magnetic, savinov2008dielectric}. Moreover, (Nd, Dy) and (La, Mn) co-doped STO showed promising results for multiferroic applications \cite{duran2005ferroelectricity, tkach2005structure, choudhury2011tuning, choudhury2013magnetization, azzoni2000magnetic, savinov2008dielectric}. 
\vspace{5mm}

Here we doped STO with two different transition metal elements which are Zr and Ni. In the stable Zr$^{4+}$ oxidation state, the empty $4d$ orbital has the possibility of inducing  the ferroelectric order in STO and thereby may enhance dielectric properties in Zr doped STO \cite{spaldin2019advances, bhatti2016synthesis, pitike2015first}. The rationale behind co-doping of Zr$^{4+}$ and Ni$^{2+}$ is that the magnetic moments of the unpaired electrons in the partially filled $3d$ orbitals in Ni$^{2+}$ oxidation state might have the potential to tune the magnetic properties of STO. We assume substitutional doping following the general rule where ions with smaller radius Ni$^{2+}$ ($r_o=0.69$ \AA) tend to substitute Ti$^{4+}$ in B site and ions with larger radius Zr$^{4+}$ ($r_o=0.80$ \AA) prefer to go in A site occupied by Sr$^{2+}$ \cite{bian2018influence}. The objective of this investigation is to synthesize a number of Zr and Ni co-doped STO materials and extensively characterize their structural, electrical and magnetic properties \cite{ahmmad2016anomalous, basith201710, soni2021effects, zhou2015microstructure}. To this aim,  undoped STO,  Zr-doped $\text{Sr}_\text{1-x}\text{Zr}_\text{x}\text{Ti}\text{O}_3$ and (Zr, Ni) co-doped $\text{Sr}_\text{1-x}\text{Zr}_\text{x}\text{Ti}_\text{1-y}\text{Ni}_\text{y}\text{O}_3$ samples were synthesized using solid state reaction technique. Based on the findings from the pertinent materials characterization techniques, we report a particular composition of this (Zr, Ni) co-doped samples with improved multiferroic properties which might have potential for multifunctional applications \cite{basith201710}.

\section{METHODOLOGY}
\subsection{Sample Preparation}
A number of samples for undoped SrTiO\textsubscript{3}, Zr-doped SrTiO\textsubscript{3} and (Zr, Ni) co-doped SrTiO\textsubscript{3} have been synthesized using the standard solid state reaction method \cite{esha20201, basith2014room}. For the starting materials, proper combinations of analytical grade SrCO\textsubscript{3} (98\% pure), TiO\textsubscript{2} (99.9\% pure), ZrO\textsubscript{2} (99.9\% pure) and NiO (99.9\% pure) were used with desired stoichiometric ratio. For the Zr-doped samples, $x = 0.02, 0.04, 0.06$ were prepared with the chemical formula $\text{Sr}_\text{1-x}\text{Zr}_\text{x}\text{Ti}\text{O}_3$. In case of (Zr, Ni) co-doped samples with chemical formula $\text{Sr}_\text{1-x}\text{Zr}_\text{x}\text{Ti}_\text{1-y}\text{Ni}_\text{y}\text{O}_3$, a series of combinations with $\text{y} = 0.05, 0.10, 0.15~\text{and}~0.20$ for fixed $\text{x}=0.04$ have been synthesized. The powder mixtures were hand milled for 6 hours using mortar and pestle to produce a homogeneous solid mixture with fine constituent particles in proximity with each other. Several droplets of polyvinyl alcohol were mixed with the samples in a steel die to facilitate binding before being subjected to uniaxial force of 20 kN in a hydraulic press to form circular disk-shaped pellets. These pellets were calcined at $800^{\circ}$C for 4 hours to promote reaction among the mixture constituents. The pre-sintered disk-shaped pellets were smashed into fine powders by hand milling in a ceramic mortar and pestle for 3 hours to expedite solid state reactions probabilities in the subsequent sintering of the samples. The crystallization temperatures of the udoped, Zr doped and (Zr, Ni) co-doped samples were measured to be $\sim$1080$^{\circ}$, $\sim$$1085^{\circ}$ and $\sim$$942^{\circ}$C respectively using a differential scanning calorimeter. The pre-sintered powder materials were pressed into circular disk shaped pellets and toroid rings by using a hydraulic press and sintered at $1250^{\circ}$C for 4 hours.

\subsection{Characterization Techniques}
To estimate the crystallization temperature of our synthesized samples, the differential scanning calorimetry was performed using a NETZSCH STA 449 F3 Jupiter simultaneous thermal analyzer. The high temperature sintering of the samples was done in a Nabertherm Muffle Furnace LT 5/14. The X-ray Diffraction (XRD) patterns for the synthesized samples were obtained from $10^{\circ}$ to $80^{\circ}$ at 35 kV accelerating voltage with an emission current of 20 mA using a Rigaku SmartLab SE multipurpose XRD system with Cu K$\alpha$ radiation ($\lambda=0.15418$ nm). The surface morphology and chemical species identification were performed with Scanning Electron Microscopy (SEM) and Energy-dispersive X-ray spectroscopy (EDX) respectively using a AVO 18 Research Scanning Electron Microscope from ZEISS.
The room temperature Raman scattering spectroscopy was performed with a Confocal Raman Microscope MonoVista CRS+ using a 532.090 nm laser. To characterize the chemical bond vibrations inside our samples, we used the KBr pellet technique in Fourier Transform Infrared (FTIR) PerkinElmer Spectrum spectrometer. The dielectric constant and resistivity of the samples were measured from complex impedance spectroscopy performed by Wayne Kerr 6500B Impedance Analyzer. For the electric hysteresis measurements, the electric polarization $P$ vs electric field $E$ loops were recorded using a Precision Multiferroic II Ferroelectric Test System from Marine India. The magnetization $M$ vs. magnetic field $H$ hysteresis loops of the samples were obtained using a vibrating sample magnetometer VSM from Quantum Design PPMS DynaCool.

\begin{table*}[ht]
    \centering
    \begin{tabular}{ c c c c c c c c c c c c c c}
     \hline
      \hline
      \multicolumn{14}{c}{XRD analysis $\text{Sr}_{1-x}\text{Zr}_x\text{Ti}_{1-y}\text{Ni}_y\text{O}_3$ samples}\\
       \hline
      \multicolumn{2}{c}{\underline{Composition}}&\multicolumn{7}{c}{\underline{Lattice Parameters}}&\multicolumn{1}{c}{\underline{Bulk Density}}&\multicolumn{1}{c}{\underline{X-ray density}}&\multicolumn{1}{c}{\underline{Porosity}}&\multicolumn{1}{c}{\underline{Crystallite Size}}&\multicolumn{1}{c}{\underline{FWHM}} \\
    \multicolumn{1}{c}{x}&\multicolumn{1}{c}{y}&\multicolumn{1}{c}{a(\AA)}&\multicolumn{1}{c}{b(\AA)}&\multicolumn{1}{c}{c(\AA)}&\multicolumn{1}{c}{c/a}&\multicolumn{1}{c}{$\alpha(^{\circ})$}&\multicolumn{1}{c}{$\beta(^{\circ})$}&\multicolumn{1}{c}{$\gamma(^{\circ})$}&\multicolumn{1}{c}{(gcm$^{-3})$}&\multicolumn{1}{c}{(gcm$^{-3})$}&\multicolumn{1}{c}{(\%)}&\multicolumn{1}{c}{(nm)}&\multicolumn{1}{c}{(110)($^{\circ}$)}\\
    \hline
    \multicolumn{1}{c}{0.00}&\multicolumn{1}{c}{0.00}&\multicolumn{1}{c}{3.908}&\multicolumn{1}{c}{3.908}&\multicolumn{1}{c}{3.908}&\multicolumn{1}{c}{1.0}&\multicolumn{1}{c}{90}&\multicolumn{1}{c}{90}&\multicolumn{1}{c}{90}&\multicolumn{1}{c}{3.99}&\multicolumn{1}{c}{5.102}&\multicolumn{1}{c}{21.7}&\multicolumn{1}{c}{108}&\multicolumn{1}{c}{0.073}\\
    \multicolumn{1}{c}{0.02}&\multicolumn{1}{c}{0.00}&\multicolumn{1}{c}{3.903}&\multicolumn{1}{c}{3.903}&\multicolumn{1}{c}{3.903}&\multicolumn{1}{c}{1.0}&\multicolumn{1}{c}{90}&\multicolumn{1}{c}{90}&\multicolumn{1}{c}{90}&\multicolumn{1}{c}{3.99}&\multicolumn{1}{c}{5.127}&\multicolumn{1}{c}{22.1}&\multicolumn{1}{c}{88}&\multicolumn{1}{c}{0.086}\\
    \multicolumn{1}{c}{0.04}&\multicolumn{1}{c}{0.00}&\multicolumn{1}{c}{3.899}&\multicolumn{1}{c}{3.899}&\multicolumn{1}{c}{3.899}&\multicolumn{1}{c}{1.0}&\multicolumn{1}{c}{90}&\multicolumn{1}{c}{90}&\multicolumn{1}{c}{90}&\multicolumn{1}{c}{3.99}&\multicolumn{1}{c}{5.145}&\multicolumn{1}{c}{22.4}&\multicolumn{1}{c}{57}&\multicolumn{1}{c}{0.135}\\
    \multicolumn{1}{c}{0.06}&\multicolumn{1}{c}{0.00}&\multicolumn{1}{c}{3.908}&\multicolumn{1}{c}{3.908}&\multicolumn{1}{c}{3.908}&\multicolumn{1}{c}{1.0}&\multicolumn{1}{c}{90}&\multicolumn{1}{c}{90}&\multicolumn{1}{c}{90}&\multicolumn{1}{c}{3.99}&\multicolumn{1}{c}{5.109}&\multicolumn{1}{c}{21.8}&\multicolumn{1}{c}{55}&\multicolumn{1}{c}{0.137}\\
    \multicolumn{1}{c}{0.04}&\multicolumn{1}{c}{0.05}&\multicolumn{1}{c}{3.905}&\multicolumn{1}{c}{3.905}&\multicolumn{1}{c}{3.905}&\multicolumn{1}{c}{1.0}&\multicolumn{1}{c}{90}&\multicolumn{1}{c}{90}&\multicolumn{1}{c}{90}&\multicolumn{1}{c}{3.99}&\multicolumn{1}{c}{5.133}&\multicolumn{1}{c}{22.2}&\multicolumn{1}{c}{100}&\multicolumn{1}{c}{0.085}\\  \multicolumn{1}{c}{0.04}&\multicolumn{1}{c}{0.10}&\multicolumn{1}{c}{3.888}&\multicolumn{1}{c}{3.888}&\multicolumn{1}{c}{3.888}&\multicolumn{1}{c}{1.0}&\multicolumn{1}{c}{90}&\multicolumn{1}{c}{90}&\multicolumn{1}{c}{90}&\multicolumn{1}{c}{3.99}&\multicolumn{1}{c}{5.218}&\multicolumn{1}{c}{23.5}&\multicolumn{1}{c}{36}&\multicolumn{1}{c}{0.193}\\
    \multicolumn{1}{c}{0.04}&\multicolumn{1}{c}{0.15}&\multicolumn{1}{c}{3.897}&\multicolumn{1}{c}{3.897}&\multicolumn{1}{c}{3.897}&\multicolumn{1}{c}{1.0}&\multicolumn{1}{c}{90}&\multicolumn{1}{c}{90}&\multicolumn{1}{c}{90}&\multicolumn{1}{c}{3.99}&\multicolumn{1}{c}{5.169}&\multicolumn{1}{c}{22.8}&\multicolumn{1}{c}{36}&\multicolumn{1}{c}{0.202}\\
    \multicolumn{1}{c}{0.04}&\multicolumn{1}{c}{0.20}&\multicolumn{1}{c}{3.894}&\multicolumn{1}{c}{3.894}&\multicolumn{1}{c}{3.894}&\multicolumn{1}{c}{1.0}&\multicolumn{1}{c}{90}&\multicolumn{1}{c}{90}&\multicolumn{1}{c}{90}&\multicolumn{1}{c}{3.99}&\multicolumn{1}{c}{5.195}&\multicolumn{1}{c}{23.1}&\multicolumn{1}{c}{49}&\multicolumn{1}{c}{0.148}\\
    \hline
    \hline
    \end{tabular}
    \caption{\label{table:XRD_Analysis} Lattice parameters a, b, c, $\alpha$, $\beta$ and $\gamma$ extracted from Rietveld refinement of XRD patterns of $\text{Sr}_\text{1-x}\text{Zr}_\text{x}\text{Ti}_\text{1-y}\text{Ni}_\text{y}\text{O}_3$ for (x, y) = (0.00 ,0.00), (0.02, 0.00), (0.04, 0.00), (0.06, 0.00), (0.04, 0.05), (0.04, 0.10), (0.04, 0.15) and (0.04, 0.20). The bulk density, X-ray density, porosity, crytallite size were derived from standard formulas. The values for FWHM were for the most intense XRD peak corresponding to (110) plane. }
\end{table*}
\section{Results and Discussion}
\subsection{\label{sec:level2}X-ray Diffraction Analysis}

\begin{figure}
	\begin{center}
		\includegraphics[scale=1]{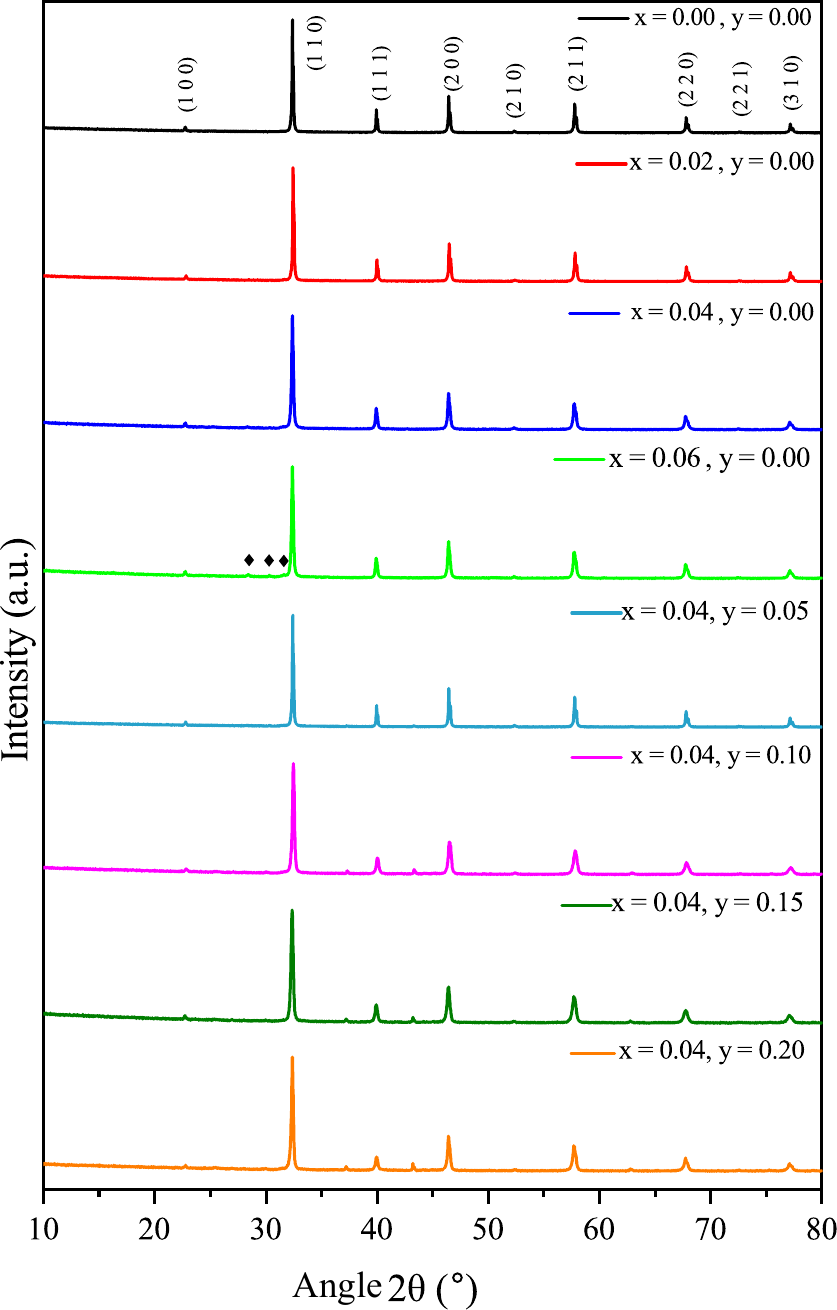}
		\caption{X-ray diffraction patterns of Sr\textsubscript{1-x}Zr\textsubscript{x}Ti\textsubscript{1-y}Ni\textsubscript{y}O\textsubscript{3} for (x, y) = (0.00, 0.00), (0.02, 0.00), (0.04, 0.00), (0.06, 0.00), (0.04, 0.05), (0.04, 0.10), (0.04, 0.15) and (0.04, 0.20). The unknown impurity phase is marked with black diamonds in case of (x, y)= (0.06, 0.00).}
		\label{XRD_Stack}
	\end{center}
\end{figure}

\begin{figure}[ht]
	\begin{center}
		\includegraphics[scale=1]{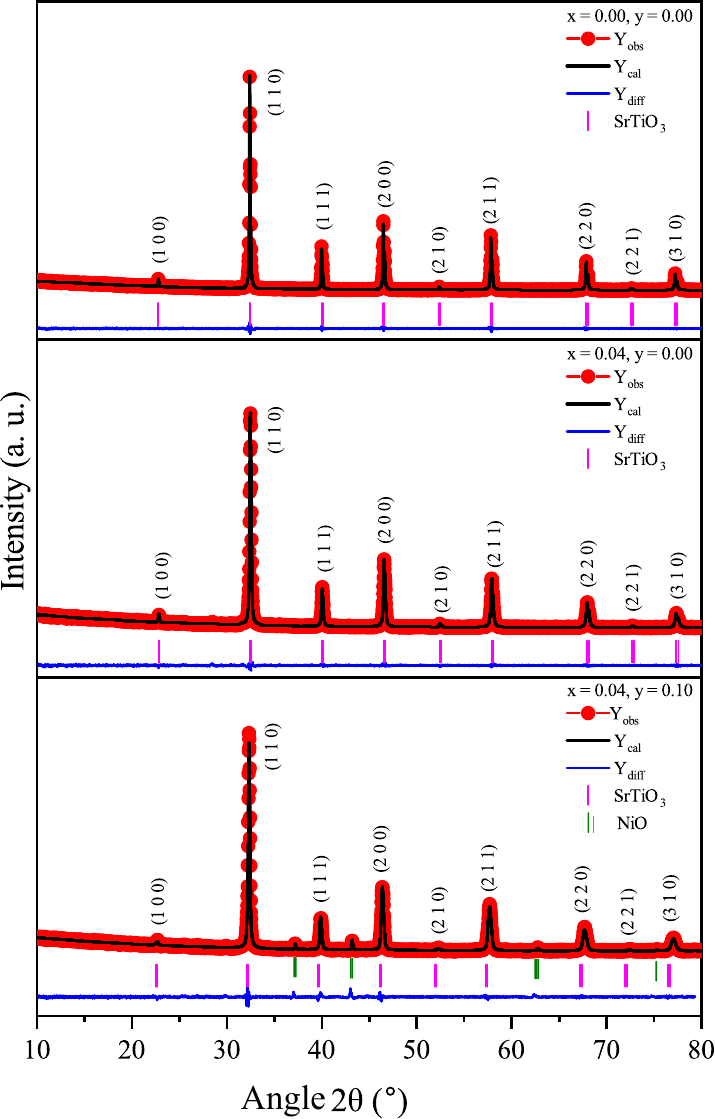}
		\caption{Simulated Rietveld
least square minimized XRD patterns superimposed on experimentally observed data of Sr\textsubscript{1-x}Zr\textsubscript{x}Ti\textsubscript{1-y}Ni\textsubscript{y}O\textsubscript{3} for (x,y) = (0.00, 0.00), (0.04, 0.00) and  (0.04, 0.10). The red solid circles are the experimental data points (Y$_\text{obs}$), the dark solid line represents calculated refined pattern Y$_\text{cal}$, the bottom blue curve Y$_\text{diff}$ shows difference between the experimental Y$_\text{obs}$ and calculated Y$_\text{cal}$ values and pink vertical lines mark the positions of Bragg peaks for cubic SrTiO\textsubscript{3} with $Pm$-$3m$ space group. The green vertical lines represent NiO phase.}
		\label{XRD_Rietveld_Refinement}
	\end{center}
\end{figure}

We have investigated the crystallographic structure, phase and purity of the synthesized samples using the XRD patterns, see Fig.~\ref{XRD_Stack}. The pure STO exhibit cubic $Pm$-$3m$ phase (space group no. 221) according to the standard JCPDS data (01-084-0443) \cite{saravanan2020effect, padmini2019investigation}. The phase purity and crystalline nature of the samples were evident from the sharp and intense diffraction peaks. In case of Zr doped $\text{Sr}_\text{1-x}\text{Zr}_\text{x}\text{Ti}\text{O}_3$ samples, extraneous peaks started to appear for $\text{x}=0.06$ indicating the presence of additional phases on top of the cubic phase. Hence, to avoid the influence of these secondary phases in (Zr, Ni) co-doped samples, we fixed Zr concentration x to be at 0.04 as $\text{Sr}_{0.96}\text{Zr}_\text{0.04}\text{Ti}_\text{1-y}\text{Ni}_\text{y}\text{O}_3$. For y$\geq0.10$, the small extraneous peaks at 37.18$^{\circ}$, 43.20$^{\circ}$, 62.75$^{\circ}$ and 75.25$^{\circ}$ were observed due excess NiO \cite{qiao2009preparation}. The lattice parameters were extracted by Rietveld refinement of the XRD profile using FullProf software. We estimated bulk density, X-ray density, porosity, crystallite size and full-width-half-maximum using standard techniques \cite{cullity1956elements}, see Table~\ref{table:XRD_Analysis}. The bulk density remained unchanged for undoped and doped STO. For Zr doped and (Zr, Ni) co-doped samples, an increasing trend in the X-ray-density was observed with Zr and Ni concentrations. These increments may have originated from the change in molecular weights due to the incorporation of the  Zr and Ni dopants in the sample. A nominal variation in porosity has been observed across the synthesized samples. As for the crystallite size, monotonic decrements have been detected with increasing Zr concentration in STO. The change in the full-width-at-half-maximum (FWHM) for (Zr, Ni) co-doped samples indicates slight distortion and disorder due to size differences of dopants and interstitial dopants respectively \cite{merupo2015structural}. From the peak shift analysis (as shown in Fig. S1 of the Supplementary Information), it is evident that the peak shift between pure STO and $\text{Sr}_{0.96}\text{Zr}_{0.04}\text{Ti}\text{O}_3$ is very nominal almost conforming with pure cubic phase. We superimposed the Rietveld refined pattern on experimentally measured XRD data for three selected samples; SrTiO\textsubscript{3}, Sr\textsubscript{0.96}Zr\textsubscript{0.04}TiO\textsubscript{3} (due to superior dielectric properties in Section \ref{sec:Dielec}) and Sr\textsubscript{0.96}Zr\textsubscript{0.04}Ti\textsubscript{0.90}Ni\textsubscript{0.10}O\textsubscript{3} (manifested superior magnetic properties as shall be seen later in Section \ref{sec:Mag}) in Fig.~\ref{XRD_Rietveld_Refinement}. The goodness of fitting parameter $\chi^2$ for the three samples were found to be 3.784, 3.590 and 3.806 which indicate simulated patterns are in good agreement with experimental observations conforming cubic structure. The auxiliary crystallographic parameters such as atomic positions in Wyckoff coordinates, relevant bond lengths and bond angles have also been extracted from the refinement (see Table S1 in the Supplementary Information).

\begin{figure}
	\begin{center}
		\includegraphics[scale=1]{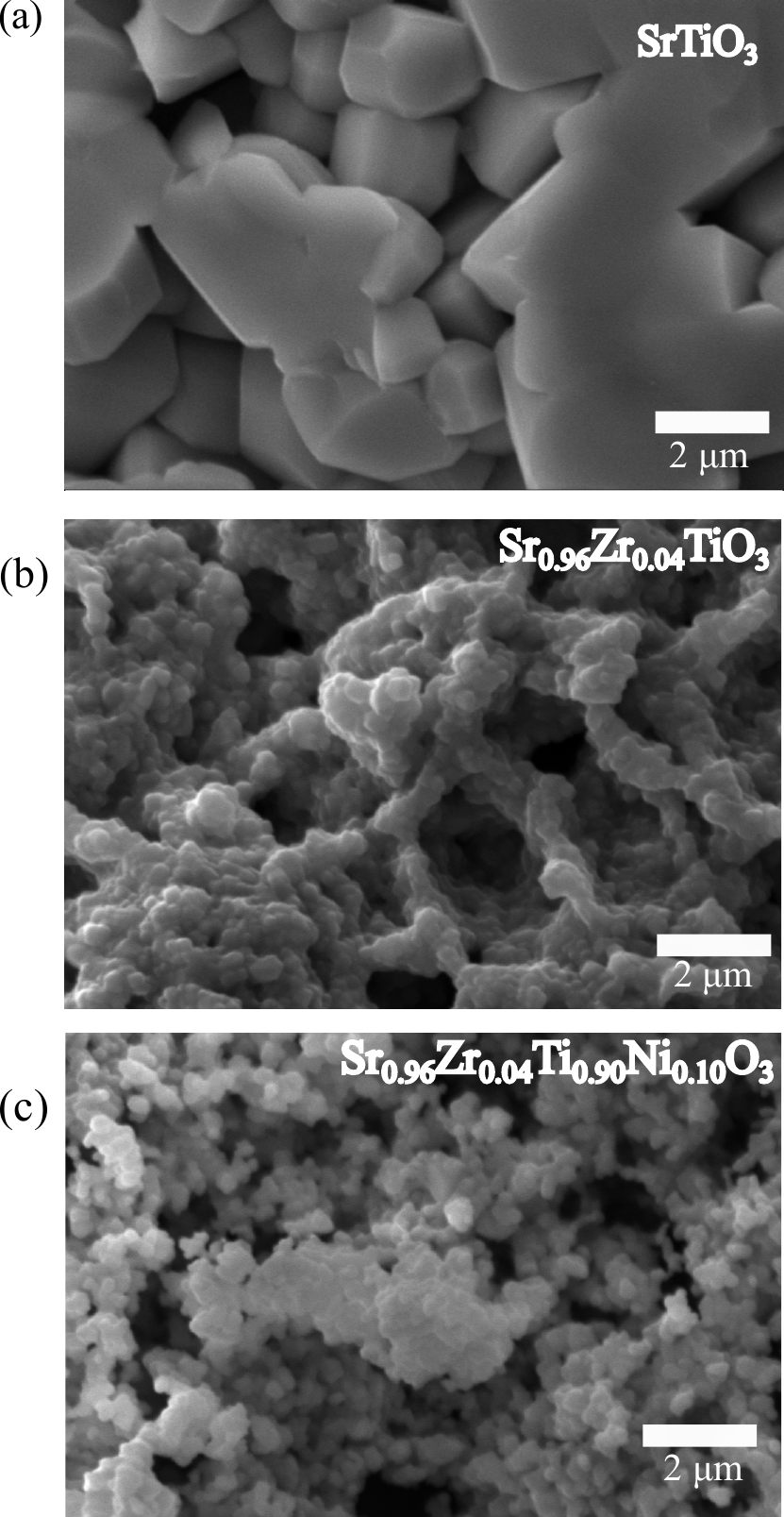}
		\caption{SEM micrographs (a) SrTiO\textsubscript{3}, (b) Sr\textsubscript{0.96}Zr\textsubscript{0.04}TiO\textsubscript{3} and (c) Sr\textsubscript{0.96}Zr\textsubscript{0.04}Ti\textsubscript{0.90}Ni\textsubscript{0.10}O\textsubscript{3} samples. }
		\label{SEM_EDX_02}
	\end{center}
\end{figure}

\subsection{\label{sec:level2}Morphological and EDX Analysis}
To understand the effect of doping on the microstructure and to perform chemical species identification of our samples, SEM micrographs and EDX spectra of the three selected samples SrTiO\textsubscript{3}, Sr\textsubscript{0.96}Zr\textsubscript{0.04}TiO\textsubscript{3} and Sr\textsubscript{0.96}Zr\textsubscript{0.04}Ti\textsubscript{0.90}Ni\textsubscript{0.10}O\textsubscript{3} have been obtained. The average grain size was estimated to be 2 $\mu\text{m}$ in case of undoped SrTiO\textsubscript{3}, see SEM micrograph in Fig.~\ref{SEM_EDX_02}(a). The estimated average grain size is comparable with the previous studies on STO in Refs. \cite{bian2018influence, rout2005study}. The detection of Sr, Ti and O peaks in the corresponding EDX spectra for undoped STO excludes the presence of unwanted chemical species in the sample (see Fig. S2(a) in the Supplementary Information). The average grain size was reduced to 0.343 $\mu\text{m}$ for Sr\textsubscript{0.96}Zr\textsubscript{0.04}TiO\textsubscript{3} sample as showed in Fig.~\ref{SEM_EDX_02}(b). This shrinkage of the gain size can be due to the presence of Zr or Ti at the grain boundaries \cite{bian2018influence}. In addition, Zr substitution prompted grain irregularity and inhomogeneity. The presence of Zr peak in addition with Sr, Ti and O peaks in the corresponding EDX spectra for Sr\textsubscript{0.96}Zr\textsubscript{0.04}TiO\textsubscript{3} corroborates its incorporation as a dopant in the sample (see Fig. S2(b) in the Supplementary Information). For (Zr, Ni) co-doped Sr\textsubscript{0.96}Zr\textsubscript{0.04}Ti\textsubscript{0.90}Ni\textsubscript{0.10}O\textsubscript{3} sample, the estimated grain size was found to be 0.341 $\mu\text{m}$ according to the SEM micrograph in Fig.~\ref{SEM_EDX_02}(c). The corresponding EDX spectra containing Zr and Ni peaks elucidates the co-doping of the STO (see Fig. S2(c) in the Supplementary Information). Moreover the atomic weights (\%) of chemical species in all aforementioned samples were compared with the corresponding theoretical values in Table S2 of the Supplementary Information.  
\subsection{\label{sec:level2}Raman Analysis}
\begin{figure}
	\begin{center}
		\includegraphics[scale=1]{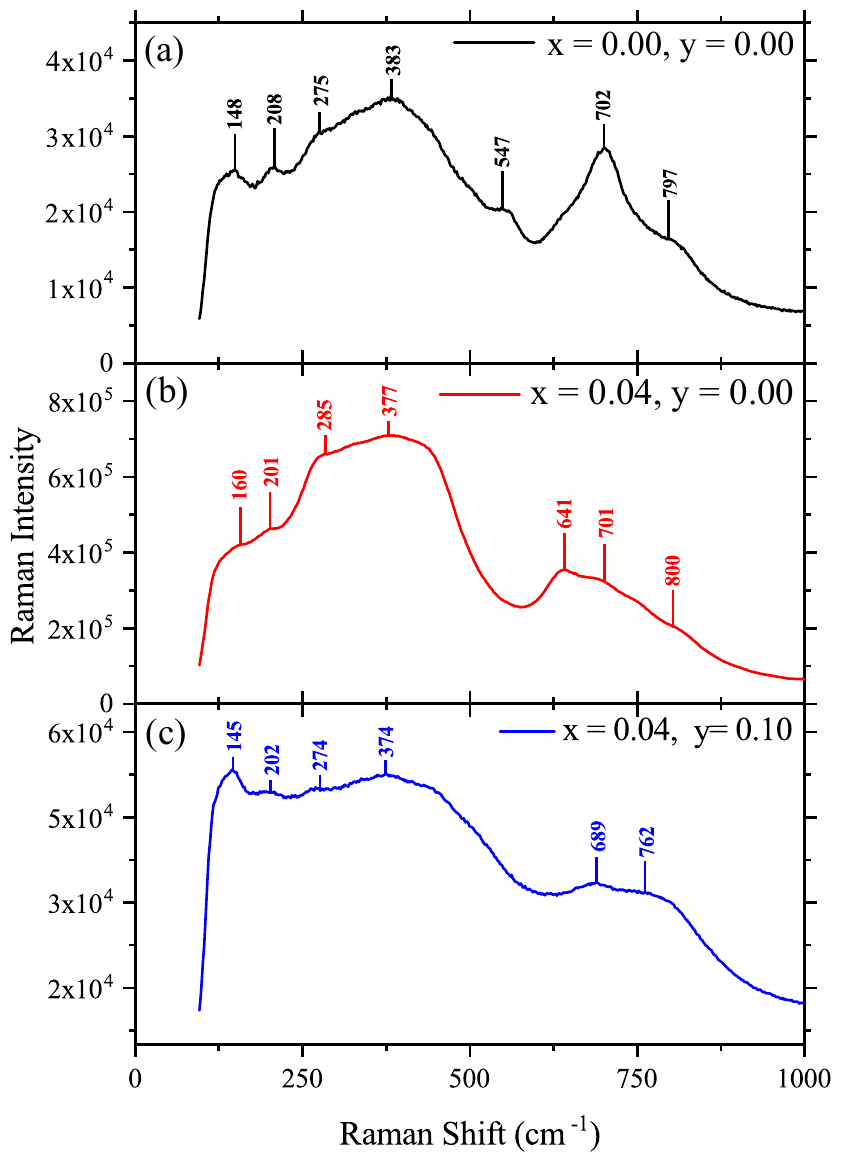}
		\caption{Raman spectra of (a) SrTiO$_3$, (b) $\text{Sr}_{0.96}\text{Zr}_{0.04}\text{Ti}\text{O}_3$ and (c) $\text{Sr}_{0.96}\text{Zr}_{0.04}\text{Ti}_{0.90}\text{Ni}_\text{0.10}\text{O}_3$ samples. }
		\label{Figure_08}
	\end{center}
\end{figure}

We performed room temperature Raman scattering spectroscopy to characterize the vibrational phonon in terms of transverse acoustic (TA), longitudinal acoustic (LA), transverse optical (TO) and longitudinal optical (LO) modes of our samples. For a three dimensional ($d=3$) unit cell with cubic symmetry, the STO contains $n=5$ atoms (one Sr, one Ti and three O) that generates $3n=15$ vibrational phonon modes; out of which the 3 low frequency acoustic modes (F$_{1u}$) are degenerate, 3 degenerate optical modes (F$_{1u}$) are Raman and infrared (IR) inactive and the rest of 9 optical modes (F$_{2u}$) are IR-active \cite{narayanan1961raman, nilsen1968raman, schaufele1967first, perry1967temperature, sirenko1999observation}. As cubic symmetry forbids first order Raman scattering, the Raman modes in cubic STO originate from second order scattering processes \cite{narayanan1961raman}. The two photon momentum conservation processes made the second order Raman scattering peaks to appear broad and continuous, see Fig.~\ref{Figure_08}. The broad intensity peaks of the Raman scattered radiation appeared in 250-500 cm$^{-1}$ and 600-800 cm$^{-1}$ wave number ranges. These Raman peaks were identified with corresponding vibrational phonon modes of the samples according to Refs. \cite{perry1967temperature, schaufele1967first} in Table S3 of the Supplementary Information. The 2TA mode appeared around 208 cm$^{-1}$, whereas the combined acoustic and optical modes such as TO$_1+$LA, LO$_1+$TA and LA+LO$_3$ contributed at 275, 383 and 702 cm$^{-1}$ respectively. The modes corresponding to two optical phonons such as TO$_1$+TO$_4$ (641 cm$^{-1}$) and LO$_3$+TO$_2$ (762 cm$^{-1}$) were there for both Zr and (Zr, Ni) co-doped STO. The presence of peaks at 148 cm$^{-1}$ (TO$_1$), 547 cm$^{-1}$ (TO$_4$) and 797 cm$^{-1}$ (LO$_4$) for undoped STO can be attributed to forbidden first order Raman scattering. This may indicate a nominal loss of inversion symmetry due to surface frozen dipoles and oxygen vacancies \cite{rabuffetti2008synthesis, tenne2007raman}. In case of doped samples, this nominal symmetry breaking can happen due to incorporation of dopants in the host STO \cite{muralidharan2014carrier}. 
\begin{figure}[ht]
	\begin{center}
		\includegraphics[scale=1]{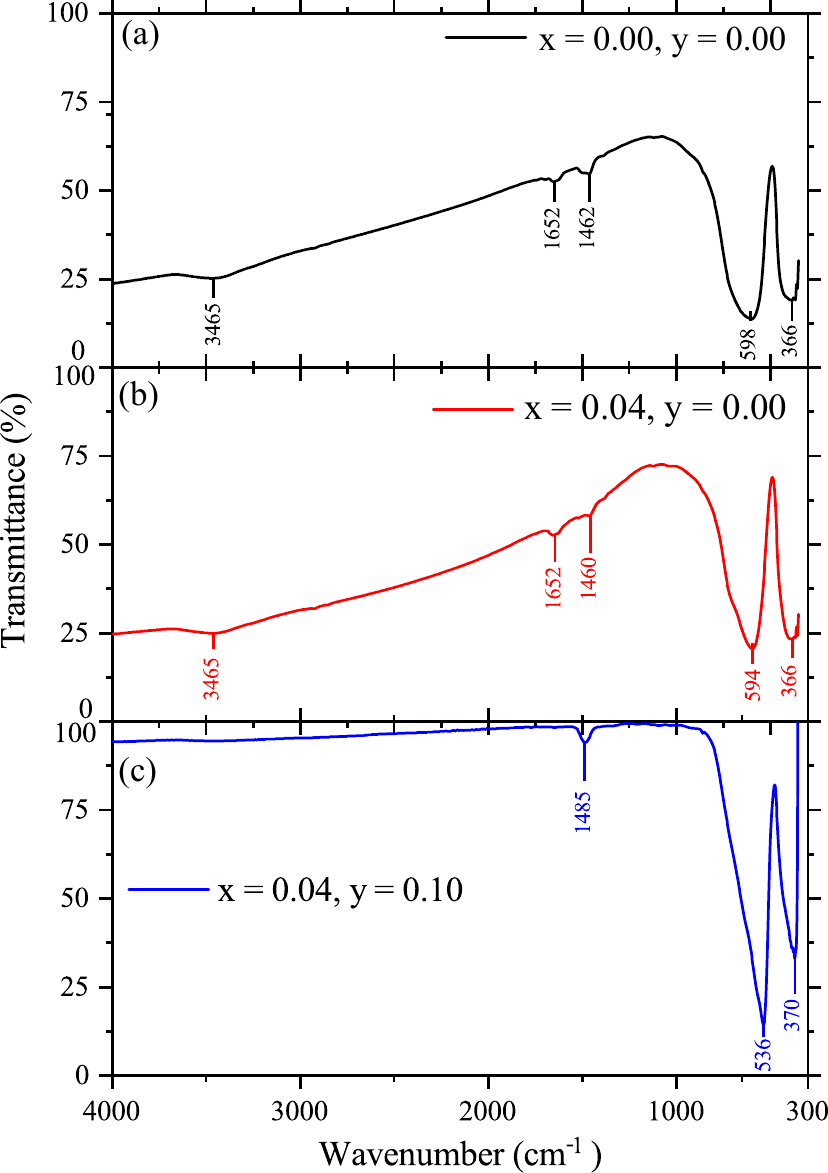}
		\caption{FTIR spectra of (a) SrTiO$_3$, (b) $\text{Sr}_{0.96}\text{Zr}_{0.04}\text{Ti}\text{O}_3$ and (c) $\text{Sr}_{0.96}\text{Zr}_{0.04}\text{Ti}_{0.90}\text{Ni}_\text{0.10}\text{O}_3$ samples. }
		\label{Figure_09}
	\end{center}
\end{figure}
\begin{figure*}[ht]
	\begin{center}
		\includegraphics[scale=1]{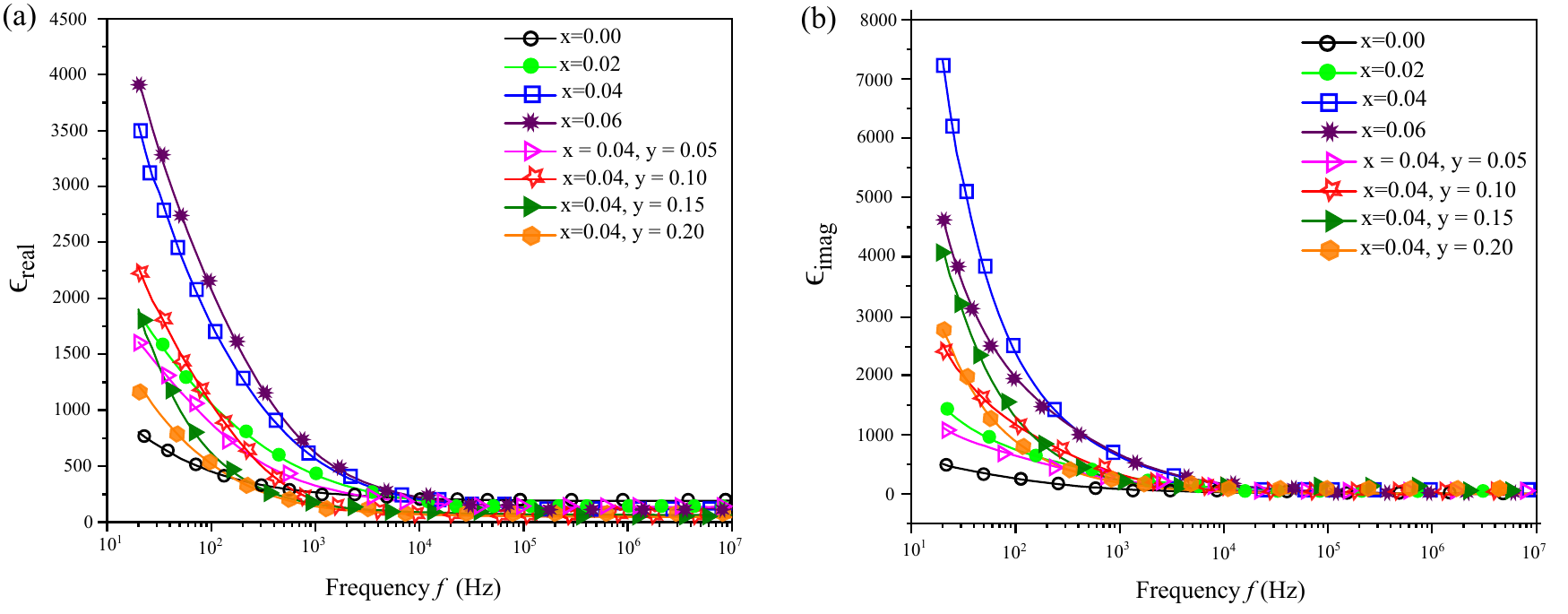}
		\caption{(a) The real $\epsilon_{\text{real}}$ vs frequency $f$ and (b) the imaginary part $\epsilon_{\text{imag}}$ vs frequency $f$ of the complex dielectric constant $\epsilon = \epsilon_{\text{real}}+i\epsilon_{\text{imag}}$ of Sr\textsubscript{1-x}Zr\textsubscript{x}Ti\textsubscript{1-y}Ni\textsubscript{y}O\textsubscript{3} for (x, y)=(0.00, 0.00), (0.02, 0.00), (0.04, 0.00), (0.06, 0.00), (0.04, 0.05), (0.04, 0.10), (0.04, 0.15) and (0.04, 0.20).}
		\label{Figure_02}
	\end{center}
\end{figure*}
\subsection{\label{sec:level2}Fourier Transform Infrared Spectroscopy}

 We also measured FTIR spectra for the undoped STO, 4\% Zr doped  $\text{Sr}_{0.96}\text{Zr}_{0.04}\text{Ti}\text{O}_3$ and (4\% Zi, 10\% Ni) co-doped $\text{Sr}_{0.96}\text{Zr}_{0.04}\text{Ti}_{0.90}\text{Ni}_{0.10}\text{O}_3$ samples at room temperature from 350 to 4000 cm$^{-1}$ and displayed in Fig.~\ref{Figure_09}. The measured FTIR absorption peaks were identified to corresponding chemical bond vibrations following Refs. \cite{xian2014photocatalytic, srilakshmi2018structural, patil2005fourier, xie2018new} in Table S4 of the Supplementary Information. The absorption peaks at 1652 cm$^{-1}$ and 3465 cm$^{-1}$ can be designated to hydroxyl –OH stretching vibration, see Fig.~\ref{Figure_09} (a)\&(b). The stemming of H\textsubscript{2}O or -OH in our sample may imply adsorption of water molecules from air surrounding the sample. The FTIR peaks at 1462, 1460  and 1485 cm$^{-1}$ for the three samples hinted deformed -OH in C-OH bond \cite{xian2014photocatalytic}. The trace of C may have its origin in SrCO\textsubscript{3} even after the calcination process. The FTIR bands in 300-600 cm$^{-1}$ represent characteristic IR absorptions due to Ti-O in STO. The peaks around $\sim$366  cm$^{-1}$ for different samples can appear from TiO\textsubscript{2} bending vibrations. Moreover the one near $\sim$598 cm$^{-1}$ can be ascribed to TiO\textsubscript{6} stretching vibration connected to Sr. We have not observed any sharp absorption peak at 500 cm$^{-1}$ in $\text{Sr}_{0.96}\text{Zr}_{0.04}\text{Ti}\text{O}_3$ which corresponds to Zr-O stretching vibrations in ZrO\textsubscript{2}, see Fig.~\ref{Figure_09}(b). This corroborates the fact that Zr$^{4+}$ ions have been incorporated in STO lattice. But Zr$^{4+}$ ions, as it replaces Sr$^{2+}$, shorten different Ti-O bond lengths to different degrees; effectively generate several very closely spaced IR absorption peaks. The combined effect of these peaks is to widen the  Sr–Ti–O absorption peak at 594 cm$^{-1}$. In case of (Zr, Ni) co-doped sample, as Ni$^{2+}$ ions replace the Ti$^{4+}$, they affect Ti-O bonds and shift the Sr–Ti–O absorption peak to 536 cm$^{-1}$ in Fig.~\ref{Figure_09}(c).

\subsection{\label{sec:Dielec}Dielectric Measurements}
The circular disk-shaped pellets were used to form parallel plate capacitors with a geometric capacitance $C_0$ giving rise to a frequency $f$ and complex dielectric constant $\epsilon = \epsilon_{\text{real}}+i\epsilon_{\text{imag}}$ ( $i=\sqrt{-1}$) dependent impedance $Z(f)=1/i2\pi fC_0 \epsilon$ which was measured with the impedance analyzer \cite{ganguly2008complex, jha2013electrical}. The real part $\epsilon_\text{real}$ of the complex dielectric constant was plotted for $\text{Sr}_\text{1-x}\text{Zr}_\text{x}\text{Ti}_\text{1-y}\text{Ni}_\text{y}\text{O}_3$ samples as a function of excitation frequency $f$ in Fig.~\ref{Figure_02}(a). The $\epsilon_\text{real}$ gradually decreases with increasing $f$ for undoped, Zr doped and (Zr, Ni) co-doped STO samples. This indicates electric dipoles inside the material struggle to synchronize and fall out of steps with high frequency electric field. The dielectric behaviour is controlled by different constituents polarizations arising from interfacial charge, space charge, oriental dipolar, ionic and electronic contributions \cite{kasap2006principles}. Owing to the fact that the $\epsilon_\text{real}$ decays rapidly beyond 1 kHz, we attribute this to interfacial and space charge polarization effects \cite{saravanan2020effect, hossain2020interrelation, muralidharan2015carrier}. This polarization may arise due to charges trapped at the interface between the sample and the electrodes, space charges at the grain boundaries, interstitial and voids. These trapped charges are sluggish in responding to the applied field in the high frequency regime beyond 1 kHz and usually modelled within the general framework of Maxwell-Wagner relaxation processes occurring inside the sample \cite{qi2018effects}. 
\begin{figure}
	\begin{center}
		\includegraphics[scale=1]{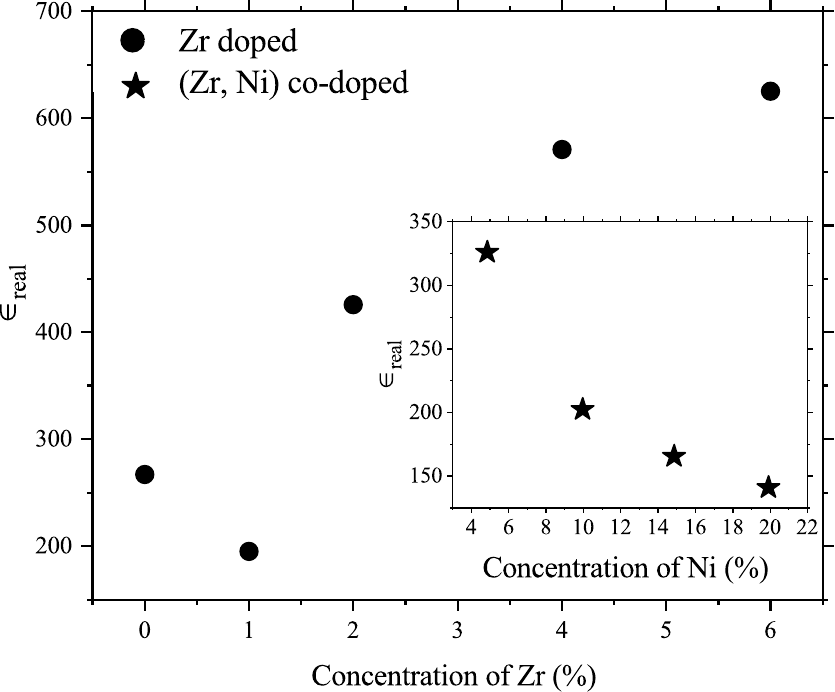}
		\caption{The $\epsilon_{\text{real}}$ of Sr\textsubscript{1-x}Zr\textsubscript{x}TiO\textsubscript{3} sample for x = 0.00, 0.02, 0.04 and 0.06. Inset: The $\epsilon_{\text{real}}$ of Sr\textsubscript{0.96}Zr\textsubscript{0.04}Ti\textsubscript{1-y}Ni\textsubscript{y}O\textsubscript{3} sample for y= 0.05, 0.10, 0.15 and 0.20. All estimations were done at $f$ = 1 kHz.}
		\label{Figure_03}
	\end{center}
\end{figure}
For the dielectric loss analysis, the imaginary part $\epsilon_\text{imag}$ of the complex $\epsilon$ was plotted as a function of frequency $f$ in Fig.~\ref{Figure_02}(b). This $\epsilon_\text{imag}$ quantifies the energy dissipation of the electric dipoles due to random collisions or phase lag during their orientation change in response to the oscillating field. The frequency response of $\epsilon_\text{imag}$ has similar trend as compared to $\epsilon_\text{real}$, i.e. it diminishes with increasing $f$ for undoped, Zr doped and (Zr, Ni) co-doped samples. This is expected as higher losses occur at low frequencies due to interfacial and space charge polarizations.  

The effect of doping on $\epsilon_\text{real}$ was analyzed in Fig.~\ref{Figure_03} for a fixed frequency of 1 kHz. For Zr doped $\text{Sr}_\text{1-x}\text{Zr}_\text{x}\text{Ti}\text{O}_3$ samples, the $\epsilon_\text{real}$ increases monotonically with increasing composition x=$0.02,~04~\text{and}~0.06$. This enhancement of dielectric constant can be attributed to more space charge accumulation due to higher oxidation state of $\text{Zr}^{4+}$ as compared to $\text{Sr}^{2+}$ in A site of STO. But in case of $\text{Sr}_{0.96}\text{Zr}_{0.04}\text{Ti}_\text{1-y}\text{Ni}_\text{y}\text{O}_3$ samples, $\epsilon_\text{real}$ decreases steadily with increasing Ni concentrations as shown in the inset of Fig.~\ref{Figure_03}. This $\epsilon_\text{real}$ reduction may imply depletion of space charges as $\text{Ni}^{2+}$ substitutes $\text{Ti}^{4+}$ ion at the B-site of STO. Moreover, the substitutional $\text{Ni}^{2+}$ can restrain the rattling of $\text{Ti}^{4+}$ ions in TiO\textsubscript{6} octahedra causing the reduction of dielectric constant \cite{bian2018influence,fu2008structure, kipkoech2005microstructural}.

\begin{figure}[ht]
	\begin{center}
		\includegraphics[scale=1]{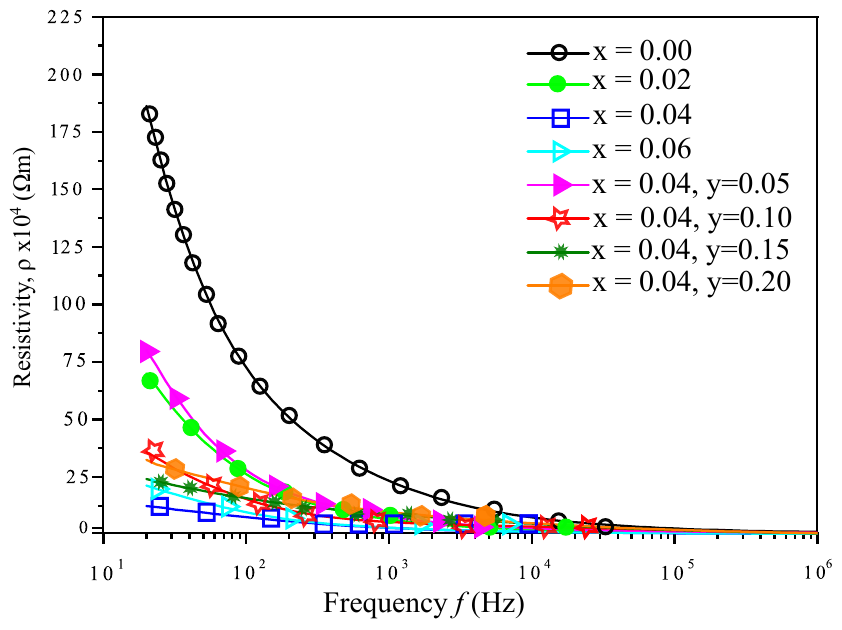}
		\caption{The resistivity $\rho$ of Sr\textsubscript{1-x}Zr\textsubscript{x}Ti\textsubscript{1-y}Ni\textsubscript{y}O\textsubscript{3} as a function of frequency $f$  for (x, y)= (0.00, 0.00), (0.02, 0.00), (0.04, 0.00), (0.06, 0.00), (0.04, 0.05), (0.04, 0.10), (0.04, 0.15) and (0.04, 0.20). }
		\label{Figure_04a}
	\end{center}
\end{figure}

\begin{figure}[ht]
	\begin{center}
		\includegraphics[scale=1]{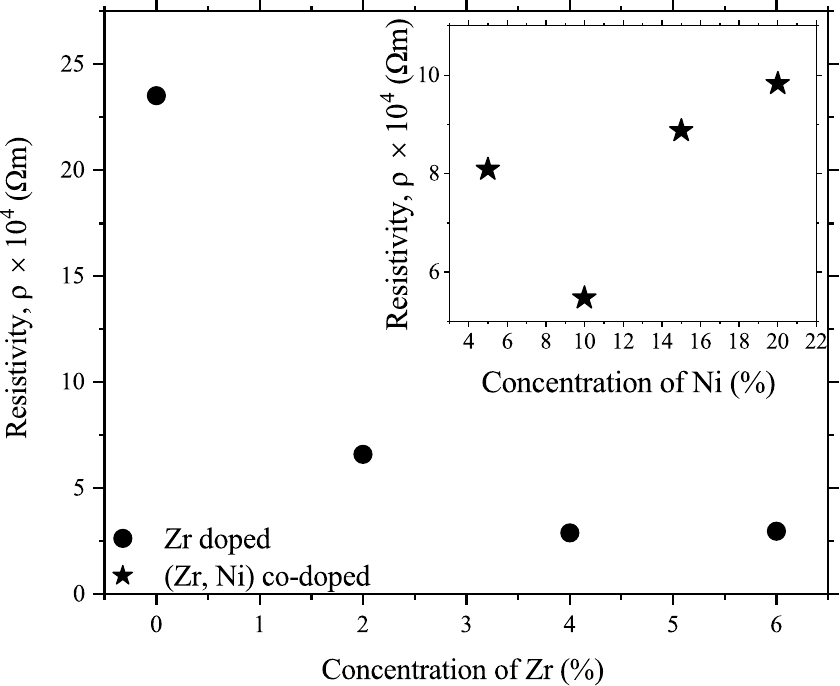}
		\caption{ The resistivity $\rho$ at a fixed frequency $f$ = 1 kHz of Sr\textsubscript{1-x}Zr\textsubscript{x}TiO\textsubscript{3} samples for x = 0.00, 0.02, 0.04 and 0.06. Inset: The $\rho$ of Sr\textsubscript{0.96}Zr\textsubscript{0.04}Ti\textsubscript{1-y}Ni\textsubscript{y}O\textsubscript{3} samples for y= 0.05, 0.10, 0.15 and 0.20.}
		\label{Figure_04b}
	\end{center}
\end{figure}
\begin{figure*}
	\begin{center}
		\includegraphics[scale=1]{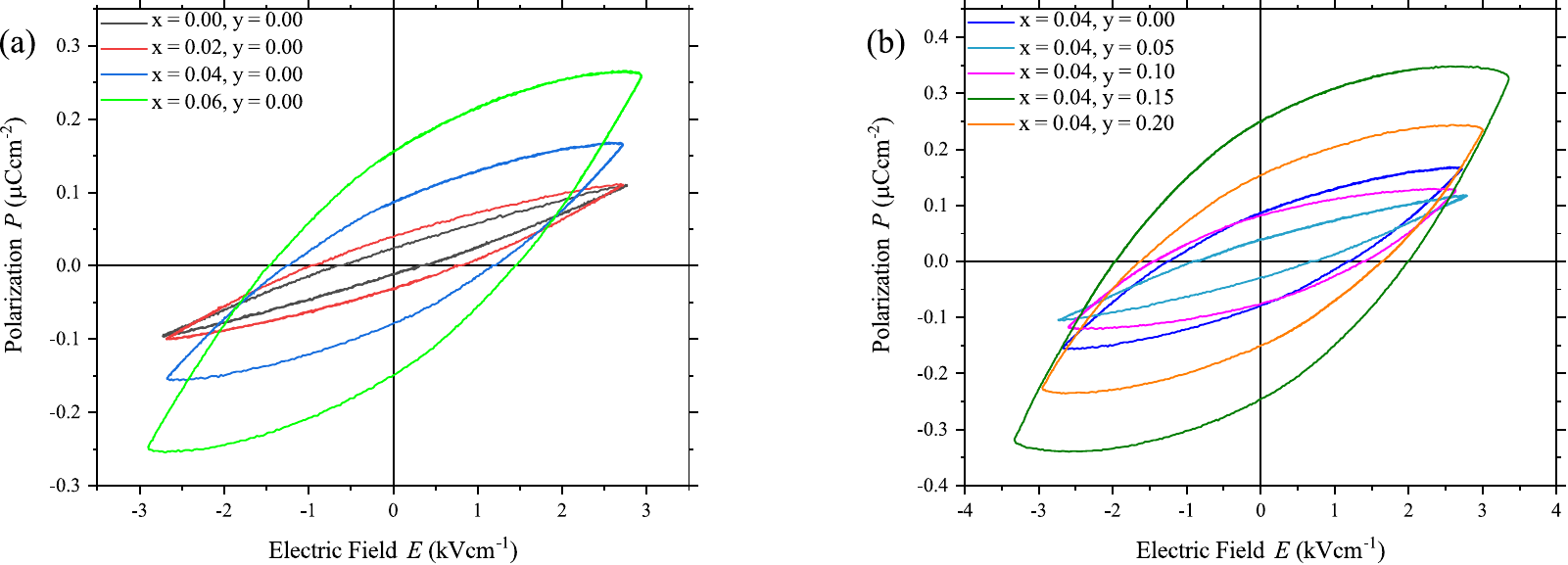}
		\caption{$P$-$E$ hysteresis loops of (a) $\text{Sr}_\text{1-x}\text{Zr}_\text{x}\text{Ti}\text{O}_3$ for $\text{x}=0.0,0.02,0.04~\text{and}~0.06$, (b) $\text{Sr}_{0.96}\text{Zr}_{0.04}\text{Ti}_\text{1-y}\text{Ni}_\text{y}\text{O}_3$ with $\text{y}=0.00, ~0.05,~0.10,~0.15~\text{and}~0.20$.  }
		\label{Figure_10}
	\end{center}
\end{figure*}

\subsection{\label{sec:level2}Resistivity Measurements}

The frequency dependant ac resistivity $\rho$ of the samples was measured from the complex impedance $Z = Z_\text{real}+iZ_\text{imag}$ by using the relation $\rho=A_\text{s}Z_\text{real}/d_\text{s}$, where $A_\text{s}$ and $d_\text{s}$ represent area and thickness of the circular disc shape pellets respectively \cite{dutta2007dielectric}. The resistivity $\rho$ decays with $f$ in Fig.~\ref{Figure_04a}. This indicates enhanced mobile charge hopping in the grain boundaries and sample-electrode interfaces \cite{trabelsi2017effect, saravanan2020effect}. The interfacial charges can produce a thin conductive layer at the sample surface effectively reducing the resistivity at high frequencies. Moreover the carrier transport in high frequency is dominated by bulk of the grains whereas grain boundary dictates the low frequency transport \cite{smari2014electric}. The carrier activation energy at the grain boundaries may fall at high frequencies resulting in enhanced charge conduction \cite{jurado2000electrical}. To analyze the effect of doping, we plot $\rho$ for undoped, Zr doped $\text{Sr}_\text{1-x}\text{Zr}_\text{x}\text{Ti}\text{O}_3$ and (Zr, Ni) co-doped $\text{Sr}_{0.96}\text{Zr}_{0.04}\text{Ti}_\text{1-y}\text{Ni}_\text{y}\text{O}_3$ samples at a fixed $f$ of 1 kHz in Fig.~\ref{Figure_04b}. The insulating characteristics of undoped STO is well captured in its high resitivity value. As we dope STO with Zr, the resistivity $\rho$ of the $\text{Sr}_{1-x}\text{Zr}_x\text{Ti}\text{O}_3$ samples decreases with increasing concentrations $x = ~0.02,~0.04$. This can be attributed to the fact that $\text{Zr}^{4+}$  acts as an n-type dopant for $\text{Sr}^{2+}$ and raises the conductivity of the samples. The Ni, as a second dopant in $\text{Sr}_{0.96}\text{Zr}_{0.04}\text{Ti}_\text{1-y}\text{Ni}_\text{y}\text{O}_3$ samples, seems to gradually increase the $\rho$ for $y=0.10,~0.15.~\text{and}~0.20$ as shown in the inset of Fig.~\ref{Figure_04b}. A number of reasons can be ascribed to this conductivity depletion. The $\text{Ni}^{2+}$ ion can act as an acceptor dopant for $\text{Ti}^{4+}$. This suppresses the n-type conductivity between the grain bulks as acceptor dopants can act as a non-conductive layer for mobile charge carriers across the grain boundaries \cite{vollman1994grain, vollmann1997grain}. Moreover, any oxygen vacancies induced by $\text{Ni}^{2+}$ can act as a charge trapping center reducing the carrier mobility.

\subsection{\label{sec:level2}Electric Hysteresis Measurements}
The room temperature electric polarization ($P$) vs electric field ($E$) hysteresis $P$-$E$ loops were measured with external triangular ac field excitation up to $\pm$3 kV$\text{cm}^{-1}$ at 50 Hz for undoped, Zr doped and (Zr, Ni) co-doped samples. The standard $P$-$E$ loop parameters such as coercive electric field $E_\text{c}$, remnant polarization $P_\text{r}$, maximum polarization $P_\text{max}$ and leakage current $I_\text{d}$ at $P_\text{max}$ were extracted and displayed in Table~\ref{table:PE_LOOP}. The undoped STO exhibits a small $P$-$E$ hysteresis loop where the $P$ is almost linear with $E$ and does not reach any saturation, see Fig.~\ref{Figure_10}(a). At $E=$3 kVcm$^{-1}$ the sample showed a $P_\text{max}=0.103$ $\mu \text{Ccm}^{-2}$, a tiny $P_\text{r}=0.017$ $\mu \text{Ccm}^{-2}$ and $E_\text{c}=0.154$ $\text{kVcm}^{-1}$. The incorporation of Zr 
seems to enhance FE behaviour in STO. For Zr doped $\text{Sr}_\text{1-x}\text{Zr}_\text{x}\text{Ti}\text{O}_3$ samples, the $P_\text{r}$ and $E_\text{c}$ values monotonically increase with the Zr concentration, see Table~\ref{table:PE_LOOP}. For (Zr, Ni) co-doped $\text{Sr}_{0.96}\text{Zr}_{0.04}\text{Ti}_\text{1-y}\text{Ni}_\text{y}\text{O}_3$ sample with a composition of y=0.05, the Ni dopants diminish the ferroelectric characteristics by narrowing down the $P$-$E$ loop ( $E_\text{c}=$0.791 $\text{kVcm}^{-1}$, $P_\text{r}=$0.039 $\mu \text{Ccm}^{-2}$ and $P_\text{max}=$0.112 $\mu \text{Ccm}^{-2}$) as compared to $\text{Sr}_{0.96}\text{Zr}_{0.04}\text{Ti}\text{O}_3$ ( $E_\text{c}=$1.216 $\text{kVcm}^{-1}$, $P_\text{r}=$0.083 $\mu \text{Ccm}^{-2}$ and $P_\text{max}=$0.162 $\mu \text{Ccm}^{-2}$), see Fig.~\ref{Figure_10}(b). The FE hysteresis existed for all (Zr, Ni) co-doped samples even for higher Ni concentrations; for example in case of $\text{Sr}_{0.96}\text{Zr}_{0.04}\text{Ti}_{0.90}\text{Ni}_\text{0.10}\text{O}_3$ sample $E_\text{c}=$1.420 $\text{kVcm}^{-1}$, $P_\text{r}=$0.079 $\mu \text{Ccm}^{-2}$ and $P_\text{max}=$0.125 $\mu \text{Ccm}^{-2}$.  
For higher doping concentration in Zr and (Zr, Ni) co-doped samples, the area of the PE loop was enlarged which indicates increased dielectric losses were present inside the sample. This is consistent with the increased leakage current in the samples as displayed in Table~\ref{table:PE_LOOP}. The origin of the leakage current can be due to carriers originated from the oxygen vacancies \cite{shibata2007xafs}. The dominance of the leakage current prevented complete saturation in polarization to occur in our samples \cite{chowdhury2017dy}.

\begin{table}[ht]
    \centering
    \begin{tabular}{ c c c c c c}
     \hline
      \hline
      \multicolumn{6}{c}{P-E Hysteresis Loop Parameters}\\
       \hline
       \multicolumn{1}{c}{$\text{x}$}&\multicolumn{1}{c}{$\text{y}$}&\multicolumn{1}{c}{$E_\text{c}$}&\multicolumn{1}{c}{$P_\text{r}$}&\multicolumn{1}{c}{$P_\text{max}$}&\multicolumn{1}{c}{$I_\text{d}$}\\
       &&\multicolumn{1}{c}{$(\text{kVcm}^{-1}$)}&\multicolumn{1}{c}{($\mu \text{Ccm}^{-2}$)}&\multicolumn{1}{c}{($\mu \text{Ccm}^{-2}$)}&\multicolumn{1}{c}{($\mu \text{Acm}^{-2}$)}\\
       \hline
      \multicolumn{1}{c}{$\text{0.00}$}&\multicolumn{1}{c}{0.00}&\multicolumn{1}{c}{0.312}&\multicolumn{1}{c}{0.011}&\multicolumn{1}{c}{0.103}&\multicolumn{1}{c}{0.073}\\
      \multicolumn{1}{c}{$\text{0.02}$}&\multicolumn{1}{c}{0.00}&\multicolumn{1}{c}{0.768}&\multicolumn{1}{c}{0.032}&\multicolumn{1}{c}{0.106}&\multicolumn{1}{c}{0.076}\\
      \multicolumn{1}{c}{$\text{0.04}$}&\multicolumn{1}{c}{0.00}&\multicolumn{1}{c}{1.216}&\multicolumn{1}{c}{0.083}&\multicolumn{1}{c}{0.162}&\multicolumn{1}{c}{0.085}\\
       \multicolumn{1}{c}{0.06}&\multicolumn{1}{c}{0.00}&\multicolumn{1}{c}{1.458}&\multicolumn{1}{c}{0.153}&\multicolumn{1}{c}{0.260}&\multicolumn{1}{c}{0.108}\\
       \multicolumn{1}{c}{0.04}&\multicolumn{1}{c}{0.05}&\multicolumn{1}{c}{0.791}&\multicolumn{1}{c}{0.039}&\multicolumn{1}{c}{0.112}&\multicolumn{1}{c}{0.077}\\
       \multicolumn{1}{c}{0.04}&\multicolumn{1}{c}{0.10}&\multicolumn{1}{c}{1.420}&\multicolumn{1}{c}{0.079}&\multicolumn{1}{c}{0.125}&\multicolumn{1}{c}{0.083}\\
       \multicolumn{1}{c}{0.04}&\multicolumn{1}{c}{0.15}&\multicolumn{1}{c}{1.983}&\multicolumn{1}{c}{0.247}&\multicolumn{1}{c}{0.344}&\multicolumn{1}{c}{0.136}\\
       \multicolumn{1}{c}{0.04}&\multicolumn{1}{c}{0.20}&\multicolumn{1}{c}{1.632}&\multicolumn{1}{c}{0.152}&\multicolumn{1}{c}{0.240}&\multicolumn{1}{c}{0.149}\\
     \hline
    \end{tabular}
    \caption{\label{table:PE_LOOP} The coercive field ($E_\text{c}$), the remnant polarization ($P_\text{r}$), the maximum polarization ($P_\text{max}$) and leakage current $I_\text{d}$ at $P_\text{max}$ of $\text{Sr}_\text{1-x}\text{Zr}_\text{x}\text{Ti}_\text{1-y}\text{Ni}_\text{y}\text{O}_3$ for different values of x and y.}
\end{table}

\begin{figure}
	\begin{center}
		\includegraphics[scale=1]{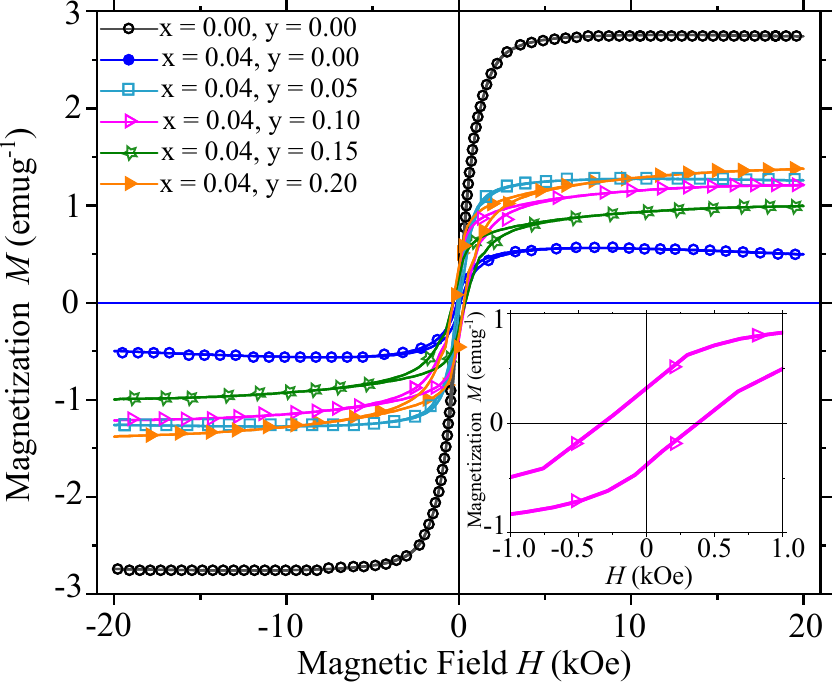}
		\caption{ Magnetic hysteresis loop of Sr\textsubscript{1-x}Zr\textsubscript{x}Ti\textsubscript{1-y}Ni\textsubscript{y}O\textsubscript{3} samples using vibrating  sample magnetometer for (x, y) = (0.00, 0.00), (0.04, 0.00), (0.04, 0.05), (0.04, 0.10), (0.04, 0.15), (0.04, 0.20). Inset: The enlarged view of the hysteresis loop for 4\% Zr and 10\% Ni co-doped sample.}
		\label{Figure_07}
	\end{center}
\end{figure}

\subsection{\label{sec:Mag}Magnetization Measurements}
We have recorded magnetization ($M$) vs. magnetic field ($H$) hysteresis loops of the as-prepared samples using a vibrating sample magnetometer applying a maximum magnetic field of $\pm$20 kOe and displayed in Fig.~\ref{Figure_07}. The inset of Fig.~\ref{Figure_07} shows an enlarged view of the hysteresis loop for the 4 \% Zr and 10 \% Ni co-doped $\text{Sr}_{0.96}\text{Zr}_{0.04}\text{Ti}_{0.90}\text{Ni}_\text{0.10}\text{O}_3$ sample. We extracted different $M$-$H$ loop parameters such as remnant magnetization ($M_\text{r}$), the coercive field ($H_\text{c}$) and the saturation magnetization ($M_\text{s}$); and the values were inserted in Table~\ref{table:MH_LOOP}. The $H_\text{c}$ was quantified following the formula $H_\text{c}=(H_\text{c1}-H_\text{c2})/2$ where $H_\text{c1}$ and $H_\text{c2}$ are the left and right coercive fields respectively \cite{basith2015tunable}. 
\begin{table}[ht]
    \centering
    \begin{tabular}{ c c c c c}
     \hline
      \hline
      \multicolumn{5}{c}{$M$-$H$ Hysteresis Loop Parameters}\\
       \hline
       \multicolumn{1}{c}{$\text{x}$}&\multicolumn{1}{c}{$\text{y}$}&\multicolumn{1}{c}{$M_\text{r}\times10^{-3}$}&\multicolumn{1}{c}{$H_\text{c}$}&\multicolumn{1}{c}{$M_\text{s}$}\\
       &&\multicolumn{1}{c}{(emug$^{-1})$}&\multicolumn{1}{c}{(Oe)}&\multicolumn{1}{c}{(emug$^{-1}$)}\\
       \hline
      \multicolumn{1}{c}{$\text{0.00}$}&\multicolumn{1}{c}{0.00}&\multicolumn{1}{c}{35}&\multicolumn{1}{c}{4}&\multicolumn{1}{c}{2.80}\\
      \multicolumn{1}{c}{$\text{0.04}$}&\multicolumn{1}{c}{0.00}&\multicolumn{1}{c}{64}&\multicolumn{1}{c}{92}&\multicolumn{1}{c}{0.50}\\
      \multicolumn{1}{c}{$\text{0.04}$}&\multicolumn{1}{c}{0.05}&\multicolumn{1}{c}{114}&\multicolumn{1}{c}{60}&\multicolumn{1}{c}{1.30}\\
       \multicolumn{1}{c}{0.04}&\multicolumn{1}{c}{0.10}&\multicolumn{1}{c}{322}&\multicolumn{1}{c}{375}&\multicolumn{1}{c}{1.27}\\
       \multicolumn{1}{c}{0.04}&\multicolumn{1}{c}{0.15}&\multicolumn{1}{c}{246}&\multicolumn{1}{c}{371}&\multicolumn{1}{c}{1.00}\\
       \multicolumn{1}{c}{0.04}&\multicolumn{1}{c}{0.20}&\multicolumn{1}{c}{365}&\multicolumn{1}{c}{308}&\multicolumn{1}{c}{0.50}\\
     \hline
    \end{tabular}
    \caption{\label{table:MH_LOOP} The remnant magnetization ($M_\text{r}$), the coercive field ($H_\text{c}$) and the saturation magnetization ($M_\text{s}$) of $\text{Sr}_\text{1-x}\text{Zr}_\text{x}\text{Ti}_\text{1-y}\text{Ni}_\text{y}\text{O}_3$ for different values of x and y.}
\end{table}

The saturated hysteresis loop with a saturation magnetization $M_\text{s}$ of 2.80 emug$^{-1}$ and coercive field of 4 Oe revealed soft FM nature of the as-synthesized STO sample. The origin of FM behaviour can be attributed to the presence oxygen vacancies $\text{V}_\text{O}^{2-}$ induced as a result of charge imbalance due to loss of Sr$^{2+}$ at high sintering temperature \cite{xu2013oxygen, verma2008resistivity, zhang2011room, crandles2010search, potzger2011defect, middey2012evidence}. Moreover the variable oxidation state of Ti ($\text{Ti}^{4+}\Leftrightarrow\text{Ti}^{3+}$) in STO can also facilitate oxygen vacancy to maintain charge equilibrium. The reduction in $M_\text{s}$ to 0.50 emug$^{-1}$ in case of $4\%$ Zr doped $\text{Sr}_{0.96}\text{Zr}_{0.04}\text{Ti}\text{O}_3$ sample can be ascribed to diamagnetic Zr$^{4+}$ ion with its empty $4d$ orbitals \cite{blundell2003magnetism}. In addition, the presence of Zr$^{4+}$ in the grain boundaries may reduce the defects at the grain surfaces that can cause suppression in $M_\text{s}$ \cite{zhang2011room}. Moreover we observed increments in $M_\text{r}$ and $H_\text{c}$ as compared to those of undoped STO which indicates reduction in softness of FM order due to Zr doping. The incorporation of Ni dopants notably increased $H_\text{c}$ and $M_\text{r}$ in (Zr, Ni) co-doped STO. In particular, for 4\% Zi and 10 \% Ni co-doped sample, the values of $H_\text{c}$, $M_\text{r}$ and $M_\text{s}$ are mentionable. For a further increment of the amount of Ni to 15 \% and 20 \%, $H_\text{c}$ and $M_\text{r}$ do not change significantly, however, $M_\text{s}$ reduced notably. Due to Zr and Ni co-doping in STO, the coercivity enhancement was very high compared to that of undoped STO which might have originated form the inflated exchange coupling between Ni$^{2+}$ ions mediated by trapped electron in the oxygen vacancy \cite{ren2007room, lei2014ferromagnetic}. Note also the variation in $H_\text{c}$ with doping concentration is not surprising as it depends on large number of factors defining the microstructure of the sample such as grain homogeneity, grain size distribution and domain wall pinning \cite{ahmmad2016large}. For the case of remnant magnetization $M_\text{r}$, we observed an order of magnitude enhancement in $\text{Sr}_{0.96}\text{Zr}_{0.04}\text{Ti}_\text{0.90}\text{Ni}_\text{0.10}\text{O}_3$ sample ($322\times10^{-3}$ emug$^{-1}$) as compared to that of undoped STO ($35\times10^{-3}$ emug$^{-1}$). Overall the presence of hysteresis loop and the remnant magnetization corroborates long range FM order in our (Zr, Ni) co-doped samples.


\section{\label{sec:level2}Conclusion}
Undoped, Zr doped and (Zr, Ni) co-doped SrTiO\textsubscript{3} samples were synthesized with varying degrees of doping concentrations and were characterized comprehensively using the appropriate techniques. We confirmed the cubic phase up to 4\% Zr doping in STO from Rietveld analysis of the powder X-ray diffraction pattern. The substitution of 4\% Zr instead of Sr in STO improved the morphological, electrical and magnetic properties. Therefore, Zr and Ni co-doped samples were prepared for this fixed \% of Zr to improve further the electrical and magnetic properties of STO. Interestingly, 4\% Zr and 10\% Ni co-doped $\text{Sr}_{0.96}\text{Zr}_{0.04}\text{Ti}_{0.90}\text{Ni}_\text{0.10}\text{O}_3$ sample demonstrated $\sim$1 and $\sim$2 orders of magnitude enhancement in remnant magnetization and coercivity respectively at room temperature. Along with a clear ferroelectric hysteresis loop we observed also a ferromagnetic hysteresis loop for this co-doped sample. We may anticipate that simultaneous existence of ferromagnetic and ferroelectric phases in this as-synthesized (Zr, Ni) co-doped $\text{Sr}_{0.96}\text{Zr}_{0.04}\text{Ti}_{0.90}\text{Ni}_\text{0.10}\text{O}_3$ sample  may open up the potentials as a multiferroic material for use in multifunctional applications.

\section*{Acknowledgments}
We gratefully acknowledge the support from Dr. Ishtiaque M. Syed for providing access to high temperature sintering facility in Semiconductor Technology Research Center, University of Dhaka.

\section*{Author contributions} 
S.A. and A. K. M. S. H. F. contributed equally. 


\textbf{Competing interests}: The authors declare no competing interests.
\clearpage

\bibliographystyle{apsrev4-1}
\bibliography{main}

\begin{thebibliography}{77}%
\makeatletter
\providecommand \@ifxundefined [1]{%
 \@ifx{#1\undefined}
}%
\providecommand \@ifnum [1]{%
 \ifnum #1\expandafter \@firstoftwo
 \else \expandafter \@secondoftwo
 \fi
}%
\providecommand \@ifx [1]{%
 \ifx #1\expandafter \@firstoftwo
 \else \expandafter \@secondoftwo
 \fi
}%
\providecommand \natexlab [1]{#1}%
\providecommand \enquote  [1]{``#1''}%
\providecommand \bibnamefont  [1]{#1}%
\providecommand \bibfnamefont [1]{#1}%
\providecommand \citenamefont [1]{#1}%
\providecommand \href@noop [0]{\@secondoftwo}%
\providecommand \href [0]{\begingroup \@sanitize@url \@href}%
\providecommand \@href[1]{\@@startlink{#1}\@@href}%
\providecommand \@@href[1]{\endgroup#1\@@endlink}%
\providecommand \@sanitize@url [0]{\catcode `\\12\catcode `\$12\catcode
  `\&12\catcode `\#12\catcode `\^12\catcode `\_12\catcode `\%12\relax}%
\providecommand \@@startlink[1]{}%
\providecommand \@@endlink[0]{}%
\providecommand \url  [0]{\begingroup\@sanitize@url \@url }%
\providecommand \@url [1]{\endgroup\@href {#1}{\urlprefix }}%
\providecommand \urlprefix  [0]{URL }%
\providecommand \Eprint [0]{\href }%
\providecommand \doibase [0]{http://dx.doi.org/}%
\providecommand \selectlanguage [0]{\@gobble}%
\providecommand \bibinfo  [0]{\@secondoftwo}%
\providecommand \bibfield  [0]{\@secondoftwo}%
\providecommand \translation [1]{[#1]}%
\providecommand \BibitemOpen [0]{}%
\providecommand \bibitemStop [0]{}%
\providecommand \bibitemNoStop [0]{.\EOS\space}%
\providecommand \EOS [0]{\spacefactor3000\relax}%
\providecommand \BibitemShut  [1]{\csname bibitem#1\endcsname}%
\let\auto@bib@innerbib\@empty
\bibitem [{\citenamefont {Robertson}(2000)}]{robertson2000band}%
  \BibitemOpen
  \bibfield  {author} {\bibinfo {author} {\bibfnamefont {J.}~\bibnamefont
  {Robertson}},\ }\href@noop {} {\bibfield  {journal} {\bibinfo  {journal}
  {Journal of Vacuum Science \& Technology B: Microelectronics and Nanometer
  Structures Processing, Measurement, and Phenomena}\ }\textbf {\bibinfo
  {volume} {18}},\ \bibinfo {pages} {1785} (\bibinfo {year}
  {2000})}\BibitemShut {NoStop}%
\bibitem [{\citenamefont {Christensen}\ \emph {et~al.}(2019)\citenamefont
  {Christensen}, \citenamefont {Trier}, \citenamefont {Niu}, \citenamefont
  {Gan}, \citenamefont {Zhang}, \citenamefont {Jespersen}, \citenamefont
  {Chen},\ and\ \citenamefont {Pryds}}]{christensen2019stimulating}%
  \BibitemOpen
  \bibfield  {author} {\bibinfo {author} {\bibfnamefont {D.~V.}\ \bibnamefont
  {Christensen}}, \bibinfo {author} {\bibfnamefont {F.}~\bibnamefont {Trier}},
  \bibinfo {author} {\bibfnamefont {W.}~\bibnamefont {Niu}}, \bibinfo {author}
  {\bibfnamefont {Y.}~\bibnamefont {Gan}}, \bibinfo {author} {\bibfnamefont
  {Y.}~\bibnamefont {Zhang}}, \bibinfo {author} {\bibfnamefont {T.~S.}\
  \bibnamefont {Jespersen}}, \bibinfo {author} {\bibfnamefont {Y.}~\bibnamefont
  {Chen}}, \ and\ \bibinfo {author} {\bibfnamefont {N.}~\bibnamefont {Pryds}},\
  }\href@noop {} {\bibfield  {journal} {\bibinfo  {journal} {Adv. Mater.
  Interfaces}\ }\textbf {\bibinfo {volume} {6}},\ \bibinfo {pages} {1900772}
  (\bibinfo {year} {2019})}\BibitemShut {NoStop}%
\bibitem [{\citenamefont {Eriksson}\ \emph {et~al.}(2003)\citenamefont
  {Eriksson}, \citenamefont {Deleniv},\ and\ \citenamefont
  {Gevorgian}}]{eriksson2003orientation}%
  \BibitemOpen
  \bibfield  {author} {\bibinfo {author} {\bibfnamefont {A.}~\bibnamefont
  {Eriksson}}, \bibinfo {author} {\bibfnamefont {A.}~\bibnamefont {Deleniv}}, \
  and\ \bibinfo {author} {\bibfnamefont {S.}~\bibnamefont {Gevorgian}},\
  }\href@noop {} {\bibfield  {journal} {\bibinfo  {journal} {Journal of Applied
  Physics}\ }\textbf {\bibinfo {volume} {93}},\ \bibinfo {pages} {2848}
  (\bibinfo {year} {2003})}\BibitemShut {NoStop}%
\bibitem [{\citenamefont {Fuchs}\ \emph {et~al.}(1999)\citenamefont {Fuchs},
  \citenamefont {Schneider}, \citenamefont {Schneider},\ and\ \citenamefont
  {Rietschel}}]{fuchs1999high}%
  \BibitemOpen
  \bibfield  {author} {\bibinfo {author} {\bibfnamefont {D.}~\bibnamefont
  {Fuchs}}, \bibinfo {author} {\bibfnamefont {C.~W.}\ \bibnamefont
  {Schneider}}, \bibinfo {author} {\bibfnamefont {R.}~\bibnamefont
  {Schneider}}, \ and\ \bibinfo {author} {\bibfnamefont {H.}~\bibnamefont
  {Rietschel}},\ }\href@noop {} {\bibfield  {journal} {\bibinfo  {journal}
  {Journal of Applied Physics}\ }\textbf {\bibinfo {volume} {85}},\ \bibinfo
  {pages} {7362} (\bibinfo {year} {1999})}\BibitemShut {NoStop}%
\bibitem [{\citenamefont {Szot}\ \emph {et~al.}(2006)\citenamefont {Szot},
  \citenamefont {Speier}, \citenamefont {Bihlmayer},\ and\ \citenamefont
  {Waser}}]{szot2006switching}%
  \BibitemOpen
  \bibfield  {author} {\bibinfo {author} {\bibfnamefont {K.}~\bibnamefont
  {Szot}}, \bibinfo {author} {\bibfnamefont {W.}~\bibnamefont {Speier}},
  \bibinfo {author} {\bibfnamefont {G.}~\bibnamefont {Bihlmayer}}, \ and\
  \bibinfo {author} {\bibfnamefont {R.}~\bibnamefont {Waser}},\ }\href@noop {}
  {\bibfield  {journal} {\bibinfo  {journal} {Nature Materials}\ }\textbf
  {\bibinfo {volume} {5}},\ \bibinfo {pages} {312} (\bibinfo {year}
  {2006})}\BibitemShut {NoStop}%
\bibitem [{\citenamefont {Hu}\ \emph {et~al.}(2004)\citenamefont {Hu},
  \citenamefont {Tan}, \citenamefont {Pan},\ and\ \citenamefont
  {Yao}}]{hu2004new}%
  \BibitemOpen
  \bibfield  {author} {\bibinfo {author} {\bibfnamefont {Y.}~\bibnamefont
  {Hu}}, \bibinfo {author} {\bibfnamefont {O.~K.}\ \bibnamefont {Tan}},
  \bibinfo {author} {\bibfnamefont {J.~S.}\ \bibnamefont {Pan}}, \ and\
  \bibinfo {author} {\bibfnamefont {X.}~\bibnamefont {Yao}},\ }\href@noop {}
  {\bibfield  {journal} {\bibinfo  {journal} {The Journal of Physical Chemistry
  B}\ }\textbf {\bibinfo {volume} {108}},\ \bibinfo {pages} {11214} (\bibinfo
  {year} {2004})}\BibitemShut {NoStop}%
\bibitem [{\citenamefont {Kajale}\ \emph {et~al.}(2012)\citenamefont {Kajale},
  \citenamefont {Patil}, \citenamefont {Gaikwad}, \citenamefont {Shinde},
  \citenamefont {Chavan}, \citenamefont {Pawar}, \citenamefont {Shirsath},\
  and\ \citenamefont {Jain}}]{kajale2012synthesis}%
  \BibitemOpen
  \bibfield  {author} {\bibinfo {author} {\bibfnamefont {D.~D.}\ \bibnamefont
  {Kajale}}, \bibinfo {author} {\bibfnamefont {G.~E.}\ \bibnamefont {Patil}},
  \bibinfo {author} {\bibfnamefont {V.~B.}\ \bibnamefont {Gaikwad}}, \bibinfo
  {author} {\bibfnamefont {S.~D.}\ \bibnamefont {Shinde}}, \bibinfo {author}
  {\bibfnamefont {D.~N.}\ \bibnamefont {Chavan}}, \bibinfo {author}
  {\bibfnamefont {N.~K.}\ \bibnamefont {Pawar}}, \bibinfo {author}
  {\bibfnamefont {S.~R.}\ \bibnamefont {Shirsath}}, \ and\ \bibinfo {author}
  {\bibfnamefont {G.~H.}\ \bibnamefont {Jain}},\ }\href@noop {} {\bibfield
  {journal} {\bibinfo  {journal} {International Journal on Smart Sensing \&
  Intelligent Systems}\ }\textbf {\bibinfo {volume} {5}} (\bibinfo {year}
  {2012})}\BibitemShut {NoStop}%
\bibitem [{\citenamefont {Sakata}\ \emph {et~al.}(2016)\citenamefont {Sakata},
  \citenamefont {Miyoshi}, \citenamefont {Maeda}, \citenamefont {Ishikiriyama},
  \citenamefont {Yamazaki}, \citenamefont {Imamura}, \citenamefont {Ham},
  \citenamefont {Hisatomi}, \citenamefont {Kubota}, \citenamefont {Yamakata}
  \emph {et~al.}}]{sakata2016photocatalytic}%
  \BibitemOpen
  \bibfield  {author} {\bibinfo {author} {\bibfnamefont {Y.}~\bibnamefont
  {Sakata}}, \bibinfo {author} {\bibfnamefont {Y.}~\bibnamefont {Miyoshi}},
  \bibinfo {author} {\bibfnamefont {T.}~\bibnamefont {Maeda}}, \bibinfo
  {author} {\bibfnamefont {K.}~\bibnamefont {Ishikiriyama}}, \bibinfo {author}
  {\bibfnamefont {Y.}~\bibnamefont {Yamazaki}}, \bibinfo {author}
  {\bibfnamefont {H.}~\bibnamefont {Imamura}}, \bibinfo {author} {\bibfnamefont
  {Y.}~\bibnamefont {Ham}}, \bibinfo {author} {\bibfnamefont {T.}~\bibnamefont
  {Hisatomi}}, \bibinfo {author} {\bibfnamefont {J.}~\bibnamefont {Kubota}},
  \bibinfo {author} {\bibfnamefont {A.}~\bibnamefont {Yamakata}},  \emph
  {et~al.},\ }\href@noop {} {\bibfield  {journal} {\bibinfo  {journal} {Applied
  Catalysis A: General}\ }\textbf {\bibinfo {volume} {521}},\ \bibinfo {pages}
  {227} (\bibinfo {year} {2016})}\BibitemShut {NoStop}%
\bibitem [{\citenamefont {Li}\ \emph {et~al.}(2004)\citenamefont {Li},
  \citenamefont {Soh},\ and\ \citenamefont {Wu}}]{li2004formability}%
  \BibitemOpen
  \bibfield  {author} {\bibinfo {author} {\bibfnamefont {C.}~\bibnamefont
  {Li}}, \bibinfo {author} {\bibfnamefont {K.~C.~K.}\ \bibnamefont {Soh}}, \
  and\ \bibinfo {author} {\bibfnamefont {P.}~\bibnamefont {Wu}},\ }\href@noop
  {} {\bibfield  {journal} {\bibinfo  {journal} {Journal of Alloys and
  Compounds}\ }\textbf {\bibinfo {volume} {372}},\ \bibinfo {pages} {40}
  (\bibinfo {year} {2004})}\BibitemShut {NoStop}%
\bibitem [{\citenamefont {Spaldin}\ and\ \citenamefont
  {Ramesh}(2019)}]{spaldin2019advances}%
  \BibitemOpen
  \bibfield  {author} {\bibinfo {author} {\bibfnamefont {N.~A.}\ \bibnamefont
  {Spaldin}}\ and\ \bibinfo {author} {\bibfnamefont {R.}~\bibnamefont
  {Ramesh}},\ }\href@noop {} {\bibfield  {journal} {\bibinfo  {journal} {Nature
  materials}\ }\textbf {\bibinfo {volume} {18}},\ \bibinfo {pages} {203}
  (\bibinfo {year} {2019})}\BibitemShut {NoStop}%
\bibitem [{\citenamefont {Kustov}\ \emph {et~al.}(2020)\citenamefont {Kustov},
  \citenamefont {Liubimova},\ and\ \citenamefont {Salje}}]{kustov2020domain}%
  \BibitemOpen
  \bibfield  {author} {\bibinfo {author} {\bibfnamefont {S.}~\bibnamefont
  {Kustov}}, \bibinfo {author} {\bibfnamefont {I.}~\bibnamefont {Liubimova}}, \
  and\ \bibinfo {author} {\bibfnamefont {E.~K.~H.}\ \bibnamefont {Salje}},\
  }\href@noop {} {\bibfield  {journal} {\bibinfo  {journal} {Physical Review
  Letters}\ }\textbf {\bibinfo {volume} {124}},\ \bibinfo {pages} {016801}
  (\bibinfo {year} {2020})}\BibitemShut {NoStop}%
\bibitem [{\citenamefont {M{\"u}ller}\ and\ \citenamefont
  {Burkard}(1979)}]{muller1979srti}%
  \BibitemOpen
  \bibfield  {author} {\bibinfo {author} {\bibfnamefont {K.~A.}\ \bibnamefont
  {M{\"u}ller}}\ and\ \bibinfo {author} {\bibfnamefont {H.}~\bibnamefont
  {Burkard}},\ }\href@noop {} {\bibfield  {journal} {\bibinfo  {journal}
  {Physical Review B}\ }\textbf {\bibinfo {volume} {19}},\ \bibinfo {pages}
  {3593} (\bibinfo {year} {1979})}\BibitemShut {NoStop}%
\bibitem [{\citenamefont {Haeni}\ \emph {et~al.}(2004)\citenamefont {Haeni},
  \citenamefont {Irvin}, \citenamefont {Chang}, \citenamefont {Uecker},
  \citenamefont {Reiche}, \citenamefont {Li}, \citenamefont {Choudhury},
  \citenamefont {Tian}, \citenamefont {Hawley}, \citenamefont {Craigo} \emph
  {et~al.}}]{haeni2004room}%
  \BibitemOpen
  \bibfield  {author} {\bibinfo {author} {\bibfnamefont {J.}~\bibnamefont
  {Haeni}}, \bibinfo {author} {\bibfnamefont {P.}~\bibnamefont {Irvin}},
  \bibinfo {author} {\bibfnamefont {W.}~\bibnamefont {Chang}}, \bibinfo
  {author} {\bibfnamefont {R.}~\bibnamefont {Uecker}}, \bibinfo {author}
  {\bibfnamefont {P.}~\bibnamefont {Reiche}}, \bibinfo {author} {\bibfnamefont
  {Y.}~\bibnamefont {Li}}, \bibinfo {author} {\bibfnamefont {S.}~\bibnamefont
  {Choudhury}}, \bibinfo {author} {\bibfnamefont {W.}~\bibnamefont {Tian}},
  \bibinfo {author} {\bibfnamefont {M.}~\bibnamefont {Hawley}}, \bibinfo
  {author} {\bibfnamefont {B.}~\bibnamefont {Craigo}},  \emph {et~al.},\
  }\href@noop {} {\bibfield  {journal} {\bibinfo  {journal} {Nature}\ }\textbf
  {\bibinfo {volume} {430}},\ \bibinfo {pages} {758} (\bibinfo {year}
  {2004})}\BibitemShut {NoStop}%
\bibitem [{\citenamefont {Choudhury}\ \emph {et~al.}(2011)\citenamefont
  {Choudhury}, \citenamefont {Mukherjee}, \citenamefont {Mandal}, \citenamefont
  {Sundaresan}, \citenamefont {Waghmare}, \citenamefont {Bhattacharjee},
  \citenamefont {Mathieu}, \citenamefont {Lazor}, \citenamefont {Eriksson},
  \citenamefont {Sanyal} \emph {et~al.}}]{choudhury2011tuning}%
  \BibitemOpen
  \bibfield  {author} {\bibinfo {author} {\bibfnamefont {D.}~\bibnamefont
  {Choudhury}}, \bibinfo {author} {\bibfnamefont {S.}~\bibnamefont
  {Mukherjee}}, \bibinfo {author} {\bibfnamefont {P.}~\bibnamefont {Mandal}},
  \bibinfo {author} {\bibfnamefont {A.}~\bibnamefont {Sundaresan}}, \bibinfo
  {author} {\bibfnamefont {U.}~\bibnamefont {Waghmare}}, \bibinfo {author}
  {\bibfnamefont {S.}~\bibnamefont {Bhattacharjee}}, \bibinfo {author}
  {\bibfnamefont {R.}~\bibnamefont {Mathieu}}, \bibinfo {author} {\bibfnamefont
  {P.}~\bibnamefont {Lazor}}, \bibinfo {author} {\bibfnamefont
  {O.}~\bibnamefont {Eriksson}}, \bibinfo {author} {\bibfnamefont
  {B.}~\bibnamefont {Sanyal}},  \emph {et~al.},\ }\href@noop {} {\bibfield
  {journal} {\bibinfo  {journal} {Physical Review B}\ }\textbf {\bibinfo
  {volume} {84}},\ \bibinfo {pages} {125124} (\bibinfo {year}
  {2011})}\BibitemShut {NoStop}%
\bibitem [{\citenamefont {Yao}\ \emph {et~al.}(2011)\citenamefont {Yao},
  \citenamefont {Zhou},\ and\ \citenamefont {Ge}}]{yao2011raman}%
  \BibitemOpen
  \bibfield  {author} {\bibinfo {author} {\bibfnamefont {D.}~\bibnamefont
  {Yao}}, \bibinfo {author} {\bibfnamefont {X.}~\bibnamefont {Zhou}}, \ and\
  \bibinfo {author} {\bibfnamefont {S.}~\bibnamefont {Ge}},\ }\href@noop {}
  {\bibfield  {journal} {\bibinfo  {journal} {Applied Surface Science}\
  }\textbf {\bibinfo {volume} {257}},\ \bibinfo {pages} {9233} (\bibinfo {year}
  {2011})}\BibitemShut {NoStop}%
\bibitem [{\citenamefont {Ahmed}\ \emph {et~al.}(2019)\citenamefont {Ahmed},
  \citenamefont {Nazrul~Islam}, \citenamefont {Hossain}, \citenamefont {Kim},
  \citenamefont {Kim}, \citenamefont {Billah}, \citenamefont {Hossain},
  \citenamefont {Yamauchi},\ and\ \citenamefont {Wang}}]{ahmed2019enhancement}%
  \BibitemOpen
  \bibfield  {author} {\bibinfo {author} {\bibfnamefont {A.~J.}\ \bibnamefont
  {Ahmed}}, \bibinfo {author} {\bibfnamefont {S.~M.~K.}\ \bibnamefont
  {Nazrul~Islam}}, \bibinfo {author} {\bibfnamefont {R.}~\bibnamefont
  {Hossain}}, \bibinfo {author} {\bibfnamefont {J.}~\bibnamefont {Kim}},
  \bibinfo {author} {\bibfnamefont {M.}~\bibnamefont {Kim}}, \bibinfo {author}
  {\bibfnamefont {M.}~\bibnamefont {Billah}}, \bibinfo {author} {\bibfnamefont
  {M.~S.~A.}\ \bibnamefont {Hossain}}, \bibinfo {author} {\bibfnamefont
  {Y.}~\bibnamefont {Yamauchi}}, \ and\ \bibinfo {author} {\bibfnamefont
  {X.}~\bibnamefont {Wang}},\ }\href@noop {} {\bibfield  {journal} {\bibinfo
  {journal} {Royal Society Open Science}\ }\textbf {\bibinfo {volume} {6}},\
  \bibinfo {pages} {190870} (\bibinfo {year} {2019})}\BibitemShut {NoStop}%
\bibitem [{\citenamefont {Padmini}\ and\ \citenamefont
  {Ramachandran}(2019)}]{padmini2019investigation}%
  \BibitemOpen
  \bibfield  {author} {\bibinfo {author} {\bibfnamefont {E.}~\bibnamefont
  {Padmini}}\ and\ \bibinfo {author} {\bibfnamefont {K.}~\bibnamefont
  {Ramachandran}},\ }\href@noop {} {\bibfield  {journal} {\bibinfo  {journal}
  {Solid State Communications}\ }\textbf {\bibinfo {volume} {302}},\ \bibinfo
  {pages} {113716} (\bibinfo {year} {2019})}\BibitemShut {NoStop}%
\bibitem [{\citenamefont {Muralidharan}\ \emph {et~al.}(2014)\citenamefont
  {Muralidharan}, \citenamefont {Anbarasu}, \citenamefont {Perumal},\ and\
  \citenamefont {Sivakumar}}]{muralidharan2014carrier}%
  \BibitemOpen
  \bibfield  {author} {\bibinfo {author} {\bibfnamefont {M.}~\bibnamefont
  {Muralidharan}}, \bibinfo {author} {\bibfnamefont {V.}~\bibnamefont
  {Anbarasu}}, \bibinfo {author} {\bibfnamefont {A.~E.}\ \bibnamefont
  {Perumal}}, \ and\ \bibinfo {author} {\bibfnamefont {K.}~\bibnamefont
  {Sivakumar}},\ }\href@noop {} {\bibfield  {journal} {\bibinfo  {journal}
  {Journal of Materials Science: Materials in Electronics}\ }\textbf {\bibinfo
  {volume} {25}},\ \bibinfo {pages} {4078} (\bibinfo {year}
  {2014})}\BibitemShut {NoStop}%
\bibitem [{\citenamefont {Norton}\ \emph {et~al.}(2002)\citenamefont {Norton},
  \citenamefont {Theodoropoulou}, \citenamefont {Hebard}, \citenamefont
  {Budai}, \citenamefont {Boatner}, \citenamefont {Pearton},\ and\
  \citenamefont {Wilson}}]{norton2002properties}%
  \BibitemOpen
  \bibfield  {author} {\bibinfo {author} {\bibfnamefont {D.}~\bibnamefont
  {Norton}}, \bibinfo {author} {\bibfnamefont {N.}~\bibnamefont
  {Theodoropoulou}}, \bibinfo {author} {\bibfnamefont {A.}~\bibnamefont
  {Hebard}}, \bibinfo {author} {\bibfnamefont {J.}~\bibnamefont {Budai}},
  \bibinfo {author} {\bibfnamefont {L.}~\bibnamefont {Boatner}}, \bibinfo
  {author} {\bibfnamefont {S.}~\bibnamefont {Pearton}}, \ and\ \bibinfo
  {author} {\bibfnamefont {R.}~\bibnamefont {Wilson}},\ }\href@noop {}
  {\bibfield  {journal} {\bibinfo  {journal} {Electrochemical and Solid State
  Letters}\ }\textbf {\bibinfo {volume} {6}},\ \bibinfo {pages} {G19} (\bibinfo
  {year} {2002})}\BibitemShut {NoStop}%
\bibitem [{\citenamefont {Lee}\ \emph {et~al.}(2003)\citenamefont {Lee},
  \citenamefont {Khim}, \citenamefont {Park}, \citenamefont {Norton},
  \citenamefont {Theodoropoulou}, \citenamefont {Hebard}, \citenamefont
  {Budai}, \citenamefont {Boatner}, \citenamefont {Pearton},\ and\
  \citenamefont {Wilson}}]{lee2003magnetic}%
  \BibitemOpen
  \bibfield  {author} {\bibinfo {author} {\bibfnamefont {J.}~\bibnamefont
  {Lee}}, \bibinfo {author} {\bibfnamefont {Z.}~\bibnamefont {Khim}}, \bibinfo
  {author} {\bibfnamefont {Y.}~\bibnamefont {Park}}, \bibinfo {author}
  {\bibfnamefont {D.}~\bibnamefont {Norton}}, \bibinfo {author} {\bibfnamefont
  {N.}~\bibnamefont {Theodoropoulou}}, \bibinfo {author} {\bibfnamefont
  {A.}~\bibnamefont {Hebard}}, \bibinfo {author} {\bibfnamefont
  {J.}~\bibnamefont {Budai}}, \bibinfo {author} {\bibfnamefont
  {L.}~\bibnamefont {Boatner}}, \bibinfo {author} {\bibfnamefont
  {S.}~\bibnamefont {Pearton}}, \ and\ \bibinfo {author} {\bibfnamefont
  {R.}~\bibnamefont {Wilson}},\ }\href@noop {} {\bibfield  {journal} {\bibinfo
  {journal} {Solid-State Electronics}\ }\textbf {\bibinfo {volume} {47}},\
  \bibinfo {pages} {2225} (\bibinfo {year} {2003})}\BibitemShut {NoStop}%
\bibitem [{\citenamefont {Dur{\'a}n}\ \emph {et~al.}(2005)\citenamefont
  {Dur{\'a}n}, \citenamefont {Mart{\'\i}nez}, \citenamefont {D{\'\i}az},\ and\
  \citenamefont {Siqueiros}}]{duran2005ferroelectricity}%
  \BibitemOpen
  \bibfield  {author} {\bibinfo {author} {\bibfnamefont {A.}~\bibnamefont
  {Dur{\'a}n}}, \bibinfo {author} {\bibfnamefont {E.}~\bibnamefont
  {Mart{\'\i}nez}}, \bibinfo {author} {\bibfnamefont {J.}~\bibnamefont
  {D{\'\i}az}}, \ and\ \bibinfo {author} {\bibfnamefont {J.}~\bibnamefont
  {Siqueiros}},\ }\href@noop {} {\bibfield  {journal} {\bibinfo  {journal}
  {Journal of Applied Physics}\ }\textbf {\bibinfo {volume} {97}},\ \bibinfo
  {pages} {104109} (\bibinfo {year} {2005})}\BibitemShut {NoStop}%
\bibitem [{\citenamefont {Tkach}\ \emph {et~al.}(2005)\citenamefont {Tkach},
  \citenamefont {Vilarinho},\ and\ \citenamefont
  {Kholkin}}]{tkach2005structure}%
  \BibitemOpen
  \bibfield  {author} {\bibinfo {author} {\bibfnamefont {A.}~\bibnamefont
  {Tkach}}, \bibinfo {author} {\bibfnamefont {P.~M.}\ \bibnamefont
  {Vilarinho}}, \ and\ \bibinfo {author} {\bibfnamefont {A.~L.}\ \bibnamefont
  {Kholkin}},\ }\href@noop {} {\bibfield  {journal} {\bibinfo  {journal} {Acta
  Materialia}\ }\textbf {\bibinfo {volume} {53}},\ \bibinfo {pages} {5061}
  (\bibinfo {year} {2005})}\BibitemShut {NoStop}%
\bibitem [{\citenamefont {Choudhury}\ \emph {et~al.}(2013)\citenamefont
  {Choudhury}, \citenamefont {Pal}, \citenamefont {Sharma}, \citenamefont
  {Bhat},\ and\ \citenamefont {Sarma}}]{choudhury2013magnetization}%
  \BibitemOpen
  \bibfield  {author} {\bibinfo {author} {\bibfnamefont {D.}~\bibnamefont
  {Choudhury}}, \bibinfo {author} {\bibfnamefont {B.}~\bibnamefont {Pal}},
  \bibinfo {author} {\bibfnamefont {A.}~\bibnamefont {Sharma}}, \bibinfo
  {author} {\bibfnamefont {S.}~\bibnamefont {Bhat}}, \ and\ \bibinfo {author}
  {\bibfnamefont {D.~D.}\ \bibnamefont {Sarma}},\ }\href@noop {} {\bibfield
  {journal} {\bibinfo  {journal} {Scientific Reports}\ }\textbf {\bibinfo
  {volume} {3}},\ \bibinfo {pages} {1433} (\bibinfo {year} {2013})}\BibitemShut
  {NoStop}%
\bibitem [{\citenamefont {Azzoni}\ \emph {et~al.}(2000)\citenamefont {Azzoni},
  \citenamefont {Mozzati}, \citenamefont {Paleari}, \citenamefont {Massarotti},
  \citenamefont {Bini},\ and\ \citenamefont {Capsoni}}]{azzoni2000magnetic}%
  \BibitemOpen
  \bibfield  {author} {\bibinfo {author} {\bibfnamefont {C.}~\bibnamefont
  {Azzoni}}, \bibinfo {author} {\bibfnamefont {M.}~\bibnamefont {Mozzati}},
  \bibinfo {author} {\bibfnamefont {A.}~\bibnamefont {Paleari}}, \bibinfo
  {author} {\bibfnamefont {V.}~\bibnamefont {Massarotti}}, \bibinfo {author}
  {\bibfnamefont {M.}~\bibnamefont {Bini}}, \ and\ \bibinfo {author}
  {\bibfnamefont {D.}~\bibnamefont {Capsoni}},\ }\href@noop {} {\bibfield
  {journal} {\bibinfo  {journal} {Solid State Communications}\ }\textbf
  {\bibinfo {volume} {114}},\ \bibinfo {pages} {617} (\bibinfo {year}
  {2000})}\BibitemShut {NoStop}%
\bibitem [{\citenamefont {Savinov}\ \emph {et~al.}(2008)\citenamefont
  {Savinov}, \citenamefont {Trepakov}, \citenamefont {Syrnikov}, \citenamefont
  {{\v{Z}}elezn{\`y}}, \citenamefont {Pokorn{\`y}}, \citenamefont {Dejneka},
  \citenamefont {Jastrabik},\ and\ \citenamefont
  {Galinetto}}]{savinov2008dielectric}%
  \BibitemOpen
  \bibfield  {author} {\bibinfo {author} {\bibfnamefont {M.}~\bibnamefont
  {Savinov}}, \bibinfo {author} {\bibfnamefont {V.}~\bibnamefont {Trepakov}},
  \bibinfo {author} {\bibfnamefont {P.}~\bibnamefont {Syrnikov}}, \bibinfo
  {author} {\bibfnamefont {V.}~\bibnamefont {{\v{Z}}elezn{\`y}}}, \bibinfo
  {author} {\bibfnamefont {J.}~\bibnamefont {Pokorn{\`y}}}, \bibinfo {author}
  {\bibfnamefont {A.}~\bibnamefont {Dejneka}}, \bibinfo {author} {\bibfnamefont
  {L.}~\bibnamefont {Jastrabik}}, \ and\ \bibinfo {author} {\bibfnamefont
  {P.}~\bibnamefont {Galinetto}},\ }\href@noop {} {\bibfield  {journal}
  {\bibinfo  {journal} {Journal of Physics: Condensed Matter}\ }\textbf
  {\bibinfo {volume} {20}},\ \bibinfo {pages} {095221} (\bibinfo {year}
  {2008})}\BibitemShut {NoStop}%
\bibitem [{\citenamefont {Bhatti}\ \emph {et~al.}(2016)\citenamefont {Bhatti},
  \citenamefont {Hussain}, \citenamefont {Khan},\ and\ \citenamefont
  {Hussain}}]{bhatti2016synthesis}%
  \BibitemOpen
  \bibfield  {author} {\bibinfo {author} {\bibfnamefont {H.~S.}\ \bibnamefont
  {Bhatti}}, \bibinfo {author} {\bibfnamefont {S.~T.}\ \bibnamefont {Hussain}},
  \bibinfo {author} {\bibfnamefont {F.~A.}\ \bibnamefont {Khan}}, \ and\
  \bibinfo {author} {\bibfnamefont {S.}~\bibnamefont {Hussain}},\ }\href@noop
  {} {\bibfield  {journal} {\bibinfo  {journal} {Applied Surface Science}\
  }\textbf {\bibinfo {volume} {367}},\ \bibinfo {pages} {291} (\bibinfo {year}
  {2016})}\BibitemShut {NoStop}%
\bibitem [{\citenamefont {Pitike}\ \emph {et~al.}(2015)\citenamefont {Pitike},
  \citenamefont {Parker}, \citenamefont {Louis},\ and\ \citenamefont
  {Nakhmanson}}]{pitike2015first}%
  \BibitemOpen
  \bibfield  {author} {\bibinfo {author} {\bibfnamefont {K.~C.}\ \bibnamefont
  {Pitike}}, \bibinfo {author} {\bibfnamefont {W.~D.}\ \bibnamefont {Parker}},
  \bibinfo {author} {\bibfnamefont {L.}~\bibnamefont {Louis}}, \ and\ \bibinfo
  {author} {\bibfnamefont {S.~M.}\ \bibnamefont {Nakhmanson}},\ }\href@noop {}
  {\bibfield  {journal} {\bibinfo  {journal} {Physical Review B}\ }\textbf
  {\bibinfo {volume} {91}},\ \bibinfo {pages} {035112} (\bibinfo {year}
  {2015})}\BibitemShut {NoStop}%
\bibitem [{\citenamefont {Bian}\ \emph {et~al.}(2018)\citenamefont {Bian},
  \citenamefont {Lu}, \citenamefont {Li}, \citenamefont {Min}, \citenamefont
  {Zhu}, \citenamefont {Fu},\ and\ \citenamefont {Zhang}}]{bian2018influence}%
  \BibitemOpen
  \bibfield  {author} {\bibinfo {author} {\bibfnamefont {W.}~\bibnamefont
  {Bian}}, \bibinfo {author} {\bibfnamefont {X.}~\bibnamefont {Lu}}, \bibinfo
  {author} {\bibfnamefont {Y.}~\bibnamefont {Li}}, \bibinfo {author}
  {\bibfnamefont {C.}~\bibnamefont {Min}}, \bibinfo {author} {\bibfnamefont
  {H.}~\bibnamefont {Zhu}}, \bibinfo {author} {\bibfnamefont {Z.}~\bibnamefont
  {Fu}}, \ and\ \bibinfo {author} {\bibfnamefont {Q.}~\bibnamefont {Zhang}},\
  }\href@noop {} {\bibfield  {journal} {\bibinfo  {journal} {Journal of
  Materials Science: Materials in Electronics}\ }\textbf {\bibinfo {volume}
  {29}},\ \bibinfo {pages} {2743} (\bibinfo {year} {2018})}\BibitemShut
  {NoStop}%
\bibitem [{\citenamefont {Ahmmad}\ \emph
  {et~al.}(2016{\natexlab{a}})\citenamefont {Ahmmad}, \citenamefont {Islam},
  \citenamefont {Billah},\ and\ \citenamefont {Basith}}]{ahmmad2016anomalous}%
  \BibitemOpen
  \bibfield  {author} {\bibinfo {author} {\bibfnamefont {B.}~\bibnamefont
  {Ahmmad}}, \bibinfo {author} {\bibfnamefont {M.~Z.}\ \bibnamefont {Islam}},
  \bibinfo {author} {\bibfnamefont {A.}~\bibnamefont {Billah}}, \ and\ \bibinfo
  {author} {\bibfnamefont {M.~A.}\ \bibnamefont {Basith}},\ }\href@noop {}
  {\bibfield  {journal} {\bibinfo  {journal} {Journal of Physics D: Applied
  Physics}\ }\textbf {\bibinfo {volume} {49}},\ \bibinfo {pages} {095001}
  (\bibinfo {year} {2016}{\natexlab{a}})}\BibitemShut {NoStop}%
\bibitem [{\citenamefont {Basith}\ \emph {et~al.}(2017)\citenamefont {Basith},
  \citenamefont {Billah}, \citenamefont {Jalil}, \citenamefont {Yesmin},
  \citenamefont {Sakib}, \citenamefont {Ashik}, \citenamefont {Yousuf},
  \citenamefont {Chowdhury}, \citenamefont {Hossain}, \citenamefont {Firoz},\
  and\ \citenamefont {Ahmmad}}]{basith201710}%
  \BibitemOpen
  \bibfield  {author} {\bibinfo {author} {\bibfnamefont {M.~A.}\ \bibnamefont
  {Basith}}, \bibinfo {author} {\bibfnamefont {A.}~\bibnamefont {Billah}},
  \bibinfo {author} {\bibfnamefont {M.~A.}\ \bibnamefont {Jalil}}, \bibinfo
  {author} {\bibfnamefont {N.}~\bibnamefont {Yesmin}}, \bibinfo {author}
  {\bibfnamefont {M.~A.}\ \bibnamefont {Sakib}}, \bibinfo {author}
  {\bibfnamefont {E.~K.}\ \bibnamefont {Ashik}}, \bibinfo {author}
  {\bibfnamefont {S.~M. E.~H.}\ \bibnamefont {Yousuf}}, \bibinfo {author}
  {\bibfnamefont {S.~S.}\ \bibnamefont {Chowdhury}}, \bibinfo {author}
  {\bibfnamefont {M.~S.}\ \bibnamefont {Hossain}}, \bibinfo {author}
  {\bibfnamefont {S.~H.}\ \bibnamefont {Firoz}}, \ and\ \bibinfo {author}
  {\bibfnamefont {B.}~\bibnamefont {Ahmmad}},\ }\href@noop {} {\bibfield
  {journal} {\bibinfo  {journal} {Journal of Alloys and Compounds}\ }\textbf
  {\bibinfo {volume} {694}},\ \bibinfo {pages} {792} (\bibinfo {year}
  {2017})}\BibitemShut {NoStop}%
\bibitem [{\citenamefont {Soni}\ \emph {et~al.}(2021)\citenamefont {Soni},
  \citenamefont {Makkar},\ and\ \citenamefont {Biswas}}]{soni2021effects}%
  \BibitemOpen
  \bibfield  {author} {\bibinfo {author} {\bibfnamefont {B.}~\bibnamefont
  {Soni}}, \bibinfo {author} {\bibfnamefont {S.}~\bibnamefont {Makkar}}, \ and\
  \bibinfo {author} {\bibfnamefont {S.}~\bibnamefont {Biswas}},\ }\href@noop {}
  {\bibfield  {journal} {\bibinfo  {journal} {Materials Characterization}\
  }\textbf {\bibinfo {volume} {174}},\ \bibinfo {pages} {110990} (\bibinfo
  {year} {2021})}\BibitemShut {NoStop}%
\bibitem [{\citenamefont {Zhou}\ \emph {et~al.}(2015)\citenamefont {Zhou},
  \citenamefont {Deng}, \citenamefont {Ding}, \citenamefont {Yu}, \citenamefont
  {Yue}, \citenamefont {Yang},\ and\ \citenamefont
  {Chu}}]{zhou2015microstructure}%
  \BibitemOpen
  \bibfield  {author} {\bibinfo {author} {\bibfnamefont {W.}~\bibnamefont
  {Zhou}}, \bibinfo {author} {\bibfnamefont {H.}~\bibnamefont {Deng}}, \bibinfo
  {author} {\bibfnamefont {N.}~\bibnamefont {Ding}}, \bibinfo {author}
  {\bibfnamefont {L.}~\bibnamefont {Yu}}, \bibinfo {author} {\bibfnamefont
  {F.}~\bibnamefont {Yue}}, \bibinfo {author} {\bibfnamefont {P.}~\bibnamefont
  {Yang}}, \ and\ \bibinfo {author} {\bibfnamefont {J.}~\bibnamefont {Chu}},\
  }\href@noop {} {\bibfield  {journal} {\bibinfo  {journal} {Materials
  Characterization}\ }\textbf {\bibinfo {volume} {107}},\ \bibinfo {pages} {1}
  (\bibinfo {year} {2015})}\BibitemShut {NoStop}%
\bibitem [{\citenamefont {N}\ \emph {et~al.}(2020)\citenamefont {N},
  \citenamefont {N.~Munny}, \citenamefont {N.~I.~Khan},\ and\ \citenamefont
  {Maria}}]{esha20201}%
  \BibitemOpen
  \bibfield  {author} {\bibinfo {author} {\bibfnamefont {I.}~\bibnamefont {N},
  \bibfnamefont {Esha}}, \bibinfo {author} {\bibfnamefont {K.}~\bibnamefont
  {N.~Munny}}, \bibinfo {author} {\bibfnamefont {M.}~\bibnamefont
  {N.~I.~Khan}}, \ and\ \bibinfo {author} {\bibfnamefont {K.~H.}\ \bibnamefont
  {Maria}},\ }\href@noop {} {\bibfield  {journal} {\bibinfo  {journal} {AIP
  Advances}\ }\textbf {\bibinfo {volume} {10}},\ \bibinfo {pages} {125026}
  (\bibinfo {year} {2020})}\BibitemShut {NoStop}%
\bibitem [{\citenamefont {A.~Basith}\ \emph {et~al.}(2014)\citenamefont
  {A.~Basith}, \citenamefont {Kurni}, \citenamefont {Alam}, \citenamefont
  {Sinha},\ and\ \citenamefont {Ahmmad}}]{basith2014room}%
  \BibitemOpen
  \bibfield  {author} {\bibinfo {author} {\bibfnamefont {M.}~\bibnamefont
  {A.~Basith}}, \bibinfo {author} {\bibfnamefont {O.}~\bibnamefont {Kurni}},
  \bibinfo {author} {\bibfnamefont {M.~S.}\ \bibnamefont {Alam}}, \bibinfo
  {author} {\bibfnamefont {B.~L.}\ \bibnamefont {Sinha}}, \ and\ \bibinfo
  {author} {\bibfnamefont {B.}~\bibnamefont {Ahmmad}},\ }\href@noop {}
  {\bibfield  {journal} {\bibinfo  {journal} {Journal of Applied Physics}\
  }\textbf {\bibinfo {volume} {115}},\ \bibinfo {pages} {024102} (\bibinfo
  {year} {2014})}\BibitemShut {NoStop}%
\bibitem [{\citenamefont {Saravanan}\ \emph {et~al.}(2020)\citenamefont
  {Saravanan}, \citenamefont {Ramachandran}, \citenamefont {Gajendiran},\ and\
  \citenamefont {Padmini}}]{saravanan2020effect}%
  \BibitemOpen
  \bibfield  {author} {\bibinfo {author} {\bibfnamefont {G.}~\bibnamefont
  {Saravanan}}, \bibinfo {author} {\bibfnamefont {K.}~\bibnamefont
  {Ramachandran}}, \bibinfo {author} {\bibfnamefont {J.}~\bibnamefont
  {Gajendiran}}, \ and\ \bibinfo {author} {\bibfnamefont {E.}~\bibnamefont
  {Padmini}},\ }\href@noop {} {\bibfield  {journal} {\bibinfo  {journal}
  {Chemical Physics Letters}\ ,\ \bibinfo {pages} {137314}} (\bibinfo {year}
  {2020})}\BibitemShut {NoStop}%
\bibitem [{\citenamefont {Qiao}\ \emph {et~al.}(2009)\citenamefont {Qiao},
  \citenamefont {Wei}, \citenamefont {Yang}, \citenamefont {Zhu},\ and\
  \citenamefont {Yan}}]{qiao2009preparation}%
  \BibitemOpen
  \bibfield  {author} {\bibinfo {author} {\bibfnamefont {H.}~\bibnamefont
  {Qiao}}, \bibinfo {author} {\bibfnamefont {Z.}~\bibnamefont {Wei}}, \bibinfo
  {author} {\bibfnamefont {H.}~\bibnamefont {Yang}}, \bibinfo {author}
  {\bibfnamefont {L.}~\bibnamefont {Zhu}}, \ and\ \bibinfo {author}
  {\bibfnamefont {X.}~\bibnamefont {Yan}},\ }\href@noop {} {\bibfield
  {journal} {\bibinfo  {journal} {Journal of Nanomaterials}\ }\textbf {\bibinfo
  {volume} {2009}} (\bibinfo {year} {2009})}\BibitemShut {NoStop}%
\bibitem [{\citenamefont {Cullity}(1956)}]{cullity1956elements}%
  \BibitemOpen
  \bibfield  {author} {\bibinfo {author} {\bibfnamefont {B.~D.}\ \bibnamefont
  {Cullity}},\ }\href@noop {} {\emph {\bibinfo {title} {Elements of X-ray
  Diffraction}}}\ (\bibinfo  {publisher} {Addison-Wesley Publishing},\ \bibinfo
  {year} {1956})\BibitemShut {NoStop}%
\bibitem [{\citenamefont {Merupo}\ \emph {et~al.}(2015)\citenamefont {Merupo},
  \citenamefont {Velumani}, \citenamefont {Ordon}, \citenamefont {Errien},
  \citenamefont {Szade},\ and\ \citenamefont {Kassiba}}]{merupo2015structural}%
  \BibitemOpen
  \bibfield  {author} {\bibinfo {author} {\bibfnamefont {V.-I.}\ \bibnamefont
  {Merupo}}, \bibinfo {author} {\bibfnamefont {S.}~\bibnamefont {Velumani}},
  \bibinfo {author} {\bibfnamefont {K.}~\bibnamefont {Ordon}}, \bibinfo
  {author} {\bibfnamefont {N.}~\bibnamefont {Errien}}, \bibinfo {author}
  {\bibfnamefont {J.}~\bibnamefont {Szade}}, \ and\ \bibinfo {author}
  {\bibfnamefont {A.-H.}\ \bibnamefont {Kassiba}},\ }\href@noop {} {\bibfield
  {journal} {\bibinfo  {journal} {CrystEngComm}\ }\textbf {\bibinfo {volume}
  {17}},\ \bibinfo {pages} {3366} (\bibinfo {year} {2015})}\BibitemShut
  {NoStop}%
\bibitem [{\citenamefont {Rout}\ \emph {et~al.}(2005)\citenamefont {Rout},
  \citenamefont {Panigrahi},\ and\ \citenamefont {Bera}}]{rout2005study}%
  \BibitemOpen
  \bibfield  {author} {\bibinfo {author} {\bibfnamefont {S.~K.}\ \bibnamefont
  {Rout}}, \bibinfo {author} {\bibfnamefont {S.}~\bibnamefont {Panigrahi}}, \
  and\ \bibinfo {author} {\bibfnamefont {J.}~\bibnamefont {Bera}},\ }\href@noop
  {} {\bibfield  {journal} {\bibinfo  {journal} {Bulletin of Materials
  Science}\ }\textbf {\bibinfo {volume} {28}},\ \bibinfo {pages} {275}
  (\bibinfo {year} {2005})}\BibitemShut {NoStop}%
\bibitem [{\citenamefont {Narayanan}\ and\ \citenamefont
  {Vedam}(1961)}]{narayanan1961raman}%
  \BibitemOpen
  \bibfield  {author} {\bibinfo {author} {\bibfnamefont {P.~S.}\ \bibnamefont
  {Narayanan}}\ and\ \bibinfo {author} {\bibfnamefont {K.}~\bibnamefont
  {Vedam}},\ }\href@noop {} {\bibfield  {journal} {\bibinfo  {journal}
  {Zeitschrift f{\"u}r Physik}\ }\textbf {\bibinfo {volume} {163}},\ \bibinfo
  {pages} {158} (\bibinfo {year} {1961})}\BibitemShut {NoStop}%
\bibitem [{\citenamefont {Nilsen}\ and\ \citenamefont
  {Skinner}(1968)}]{nilsen1968raman}%
  \BibitemOpen
  \bibfield  {author} {\bibinfo {author} {\bibfnamefont {W.~G.}\ \bibnamefont
  {Nilsen}}\ and\ \bibinfo {author} {\bibfnamefont {J.~G.}\ \bibnamefont
  {Skinner}},\ }\href@noop {} {\bibfield  {journal} {\bibinfo  {journal} {The
  Journal of Chemical Physics}\ }\textbf {\bibinfo {volume} {48}},\ \bibinfo
  {pages} {2240} (\bibinfo {year} {1968})}\BibitemShut {NoStop}%
\bibitem [{\citenamefont {Schaufele}\ and\ \citenamefont
  {Weber}(1967)}]{schaufele1967first}%
  \BibitemOpen
  \bibfield  {author} {\bibinfo {author} {\bibfnamefont {R.~F.}\ \bibnamefont
  {Schaufele}}\ and\ \bibinfo {author} {\bibfnamefont {M.~J.}\ \bibnamefont
  {Weber}},\ }\href@noop {} {\bibfield  {journal} {\bibinfo  {journal} {The
  Journal of Chemical Physics}\ }\textbf {\bibinfo {volume} {46}},\ \bibinfo
  {pages} {2859} (\bibinfo {year} {1967})}\BibitemShut {NoStop}%
\bibitem [{\citenamefont {Perry}\ \emph {et~al.}(1967)\citenamefont {Perry},
  \citenamefont {Fertel},\ and\ \citenamefont
  {McNelly}}]{perry1967temperature}%
  \BibitemOpen
  \bibfield  {author} {\bibinfo {author} {\bibfnamefont {C.~H.}\ \bibnamefont
  {Perry}}, \bibinfo {author} {\bibfnamefont {J.~H.}\ \bibnamefont {Fertel}}, \
  and\ \bibinfo {author} {\bibfnamefont {T.~F.}\ \bibnamefont {McNelly}},\
  }\href@noop {} {\bibfield  {journal} {\bibinfo  {journal} {The Journal of
  Chemical Physics}\ }\textbf {\bibinfo {volume} {47}},\ \bibinfo {pages}
  {1619} (\bibinfo {year} {1967})}\BibitemShut {NoStop}%
\bibitem [{\citenamefont {Sirenko}\ \emph {et~al.}(1999)\citenamefont
  {Sirenko}, \citenamefont {Akimov}, \citenamefont {Fox}, \citenamefont
  {Clark}, \citenamefont {Li}, \citenamefont {Si},\ and\ \citenamefont
  {Xi}}]{sirenko1999observation}%
  \BibitemOpen
  \bibfield  {author} {\bibinfo {author} {\bibfnamefont {A.~A.}\ \bibnamefont
  {Sirenko}}, \bibinfo {author} {\bibfnamefont {I.~A.}\ \bibnamefont {Akimov}},
  \bibinfo {author} {\bibfnamefont {J.~R.}\ \bibnamefont {Fox}}, \bibinfo
  {author} {\bibfnamefont {A.~M.}\ \bibnamefont {Clark}}, \bibinfo {author}
  {\bibfnamefont {H.-C.}\ \bibnamefont {Li}}, \bibinfo {author} {\bibfnamefont
  {W.}~\bibnamefont {Si}}, \ and\ \bibinfo {author} {\bibfnamefont {X.~X.}\
  \bibnamefont {Xi}},\ }\href@noop {} {\bibfield  {journal} {\bibinfo
  {journal} {Physical review letters}\ }\textbf {\bibinfo {volume} {82}},\
  \bibinfo {pages} {4500} (\bibinfo {year} {1999})}\BibitemShut {NoStop}%
\bibitem [{\citenamefont {Rabuffetti}\ \emph {et~al.}(2008)\citenamefont
  {Rabuffetti}, \citenamefont {Kim}, \citenamefont {Enterkin}, \citenamefont
  {Wang}, \citenamefont {Lanier}, \citenamefont {Marks}, \citenamefont
  {Poeppelmeier},\ and\ \citenamefont {Stair}}]{rabuffetti2008synthesis}%
  \BibitemOpen
  \bibfield  {author} {\bibinfo {author} {\bibfnamefont {F.~A.}\ \bibnamefont
  {Rabuffetti}}, \bibinfo {author} {\bibfnamefont {H.-S.}\ \bibnamefont {Kim}},
  \bibinfo {author} {\bibfnamefont {J.~A.}\ \bibnamefont {Enterkin}}, \bibinfo
  {author} {\bibfnamefont {Y.}~\bibnamefont {Wang}}, \bibinfo {author}
  {\bibfnamefont {C.~H.}\ \bibnamefont {Lanier}}, \bibinfo {author}
  {\bibfnamefont {L.~D.}\ \bibnamefont {Marks}}, \bibinfo {author}
  {\bibfnamefont {K.~R.}\ \bibnamefont {Poeppelmeier}}, \ and\ \bibinfo
  {author} {\bibfnamefont {P.~C.}\ \bibnamefont {Stair}},\ }\href@noop {}
  {\bibfield  {journal} {\bibinfo  {journal} {Chemistry of Materials}\ }\textbf
  {\bibinfo {volume} {20}},\ \bibinfo {pages} {5628} (\bibinfo {year}
  {2008})}\BibitemShut {NoStop}%
\bibitem [{\citenamefont {Tenne}\ \emph {et~al.}(2007)\citenamefont {Tenne},
  \citenamefont {Gonenli}, \citenamefont {Soukiassian}, \citenamefont {Schlom},
  \citenamefont {Nakhmanson}, \citenamefont {Rabe},\ and\ \citenamefont
  {Xi}}]{tenne2007raman}%
  \BibitemOpen
  \bibfield  {author} {\bibinfo {author} {\bibfnamefont {D.~A.}\ \bibnamefont
  {Tenne}}, \bibinfo {author} {\bibfnamefont {I.~E.}\ \bibnamefont {Gonenli}},
  \bibinfo {author} {\bibfnamefont {A.}~\bibnamefont {Soukiassian}}, \bibinfo
  {author} {\bibfnamefont {D.~G.}\ \bibnamefont {Schlom}}, \bibinfo {author}
  {\bibfnamefont {S.~M.}\ \bibnamefont {Nakhmanson}}, \bibinfo {author}
  {\bibfnamefont {K.~M.}\ \bibnamefont {Rabe}}, \ and\ \bibinfo {author}
  {\bibfnamefont {X.~X.}\ \bibnamefont {Xi}},\ }\href@noop {} {\bibfield
  {journal} {\bibinfo  {journal} {Physical Review B}\ }\textbf {\bibinfo
  {volume} {76}},\ \bibinfo {pages} {024303} (\bibinfo {year}
  {2007})}\BibitemShut {NoStop}%
\bibitem [{\citenamefont {Xian}\ \emph {et~al.}(2014)\citenamefont {Xian},
  \citenamefont {Yang}, \citenamefont {Di}, \citenamefont {Ma}, \citenamefont
  {Zhang},\ and\ \citenamefont {Dai}}]{xian2014photocatalytic}%
  \BibitemOpen
  \bibfield  {author} {\bibinfo {author} {\bibfnamefont {T.}~\bibnamefont
  {Xian}}, \bibinfo {author} {\bibfnamefont {H.}~\bibnamefont {Yang}}, \bibinfo
  {author} {\bibfnamefont {L.}~\bibnamefont {Di}}, \bibinfo {author}
  {\bibfnamefont {J.}~\bibnamefont {Ma}}, \bibinfo {author} {\bibfnamefont
  {H.}~\bibnamefont {Zhang}}, \ and\ \bibinfo {author} {\bibfnamefont
  {J.}~\bibnamefont {Dai}},\ }\href@noop {} {\bibfield  {journal} {\bibinfo
  {journal} {Nanoscale Research Letters}\ }\textbf {\bibinfo {volume} {9}},\
  \bibinfo {pages} {1} (\bibinfo {year} {2014})}\BibitemShut {NoStop}%
\bibitem [{\citenamefont {Srilakshmi}\ \emph {et~al.}(2018)\citenamefont
  {Srilakshmi}, \citenamefont {Saraf},\ and\ \citenamefont
  {Shivakumara}}]{srilakshmi2018structural}%
  \BibitemOpen
  \bibfield  {author} {\bibinfo {author} {\bibfnamefont {C.}~\bibnamefont
  {Srilakshmi}}, \bibinfo {author} {\bibfnamefont {R.}~\bibnamefont {Saraf}}, \
  and\ \bibinfo {author} {\bibfnamefont {C.}~\bibnamefont {Shivakumara}},\
  }\href@noop {} {\bibfield  {journal} {\bibinfo  {journal} {ACS Omega}\
  }\textbf {\bibinfo {volume} {3}},\ \bibinfo {pages} {10503} (\bibinfo {year}
  {2018})}\BibitemShut {NoStop}%
\bibitem [{\citenamefont {Patil}\ \emph {et~al.}(2005)\citenamefont {Patil},
  \citenamefont {Shah}, \citenamefont {Blum},\ and\ \citenamefont
  {Rahaman}}]{patil2005fourier}%
  \BibitemOpen
  \bibfield  {author} {\bibinfo {author} {\bibfnamefont {S.~K.}\ \bibnamefont
  {Patil}}, \bibinfo {author} {\bibfnamefont {N.}~\bibnamefont {Shah}},
  \bibinfo {author} {\bibfnamefont {F.~D.}\ \bibnamefont {Blum}}, \ and\
  \bibinfo {author} {\bibfnamefont {M.~N.}\ \bibnamefont {Rahaman}},\
  }\href@noop {} {\bibfield  {journal} {\bibinfo  {journal} {Journal of
  Materials Research}\ }\textbf {\bibinfo {volume} {20}},\ \bibinfo {pages}
  {3312} (\bibinfo {year} {2005})}\BibitemShut {NoStop}%
\bibitem [{\citenamefont {Xie}\ \emph {et~al.}(2018)\citenamefont {Xie},
  \citenamefont {Wang}, \citenamefont {Liu},\ and\ \citenamefont
  {Xu}}]{xie2018new}%
  \BibitemOpen
  \bibfield  {author} {\bibinfo {author} {\bibfnamefont {T.}~\bibnamefont
  {Xie}}, \bibinfo {author} {\bibfnamefont {Y.}~\bibnamefont {Wang}}, \bibinfo
  {author} {\bibfnamefont {C.}~\bibnamefont {Liu}}, \ and\ \bibinfo {author}
  {\bibfnamefont {L.}~\bibnamefont {Xu}},\ }\href@noop {} {\bibfield  {journal}
  {\bibinfo  {journal} {Materials}\ }\textbf {\bibinfo {volume} {11}},\
  \bibinfo {pages} {646} (\bibinfo {year} {2018})}\BibitemShut {NoStop}%
\bibitem [{\citenamefont {Ganguly}\ \emph {et~al.}(2008)\citenamefont
  {Ganguly}, \citenamefont {Jha},\ and\ \citenamefont
  {Deori}}]{ganguly2008complex}%
  \BibitemOpen
  \bibfield  {author} {\bibinfo {author} {\bibfnamefont {P.}~\bibnamefont
  {Ganguly}}, \bibinfo {author} {\bibfnamefont {A.}~\bibnamefont {Jha}}, \ and\
  \bibinfo {author} {\bibfnamefont {K.}~\bibnamefont {Deori}},\ }\href@noop {}
  {\bibfield  {journal} {\bibinfo  {journal} {Solid State Communications}\
  }\textbf {\bibinfo {volume} {146}},\ \bibinfo {pages} {472} (\bibinfo {year}
  {2008})}\BibitemShut {NoStop}%
\bibitem [{\citenamefont {Priyanka}\ and\ \citenamefont
  {Jha}(2013)}]{jha2013electrical}%
  \BibitemOpen
  \bibfield  {author} {\bibinfo {author} {\bibnamefont {Priyanka}}\ and\
  \bibinfo {author} {\bibfnamefont {A.}~\bibnamefont {Jha}},\ }\href@noop {}
  {\bibfield  {journal} {\bibinfo  {journal} {Bulletin of Materials Science}\
  }\textbf {\bibinfo {volume} {36}},\ \bibinfo {pages} {135} (\bibinfo {year}
  {2013})}\BibitemShut {NoStop}%
\bibitem [{\citenamefont {Kasap}(2006)}]{kasap2006principles}%
  \BibitemOpen
  \bibfield  {author} {\bibinfo {author} {\bibfnamefont {S.~O.}\ \bibnamefont
  {Kasap}},\ }\href@noop {} {\emph {\bibinfo {title} {Principles of Electronic
  Materials and Devices}}}\ (\bibinfo  {publisher} {Tata McGraw-Hill},\
  \bibinfo {year} {2006})\BibitemShut {NoStop}%
\bibitem [{\citenamefont {Hossain}\ \emph {et~al.}(2020)\citenamefont
  {Hossain}, \citenamefont {Esha}, \citenamefont {Elius}, \citenamefont
  {Khan},\ and\ \citenamefont {Maria}}]{hossain2020interrelation}%
  \BibitemOpen
  \bibfield  {author} {\bibinfo {author} {\bibfnamefont {A.}~\bibnamefont
  {Hossain}}, \bibinfo {author} {\bibfnamefont {I.~N.}\ \bibnamefont {Esha}},
  \bibinfo {author} {\bibfnamefont {I.~B.}\ \bibnamefont {Elius}}, \bibinfo
  {author} {\bibfnamefont {M.~N.~I.}\ \bibnamefont {Khan}}, \ and\ \bibinfo
  {author} {\bibfnamefont {K.~H.}\ \bibnamefont {Maria}},\ }\href@noop {}
  {\bibfield  {journal} {\bibinfo  {journal} {Journal of Materials Science:
  Materials in Electronics}\ ,\ \bibinfo {pages} {1}} (\bibinfo {year}
  {2020})}\BibitemShut {NoStop}%
\bibitem [{\citenamefont {Muralidharan}\ \emph {et~al.}(2015)\citenamefont
  {Muralidharan}, \citenamefont {Anbarasu}, \citenamefont {Perumal},\ and\
  \citenamefont {Sivakumar}}]{muralidharan2015carrier}%
  \BibitemOpen
  \bibfield  {author} {\bibinfo {author} {\bibfnamefont {M.}~\bibnamefont
  {Muralidharan}}, \bibinfo {author} {\bibfnamefont {V.}~\bibnamefont
  {Anbarasu}}, \bibinfo {author} {\bibfnamefont {A.~E.}\ \bibnamefont
  {Perumal}}, \ and\ \bibinfo {author} {\bibfnamefont {K.}~\bibnamefont
  {Sivakumar}},\ }\href@noop {} {\bibfield  {journal} {\bibinfo  {journal}
  {Journal of Materials Science: Materials in Electronics}\ }\textbf {\bibinfo
  {volume} {26}},\ \bibinfo {pages} {6352} (\bibinfo {year}
  {2015})}\BibitemShut {NoStop}%
\bibitem [{\citenamefont {Qi}\ \emph {et~al.}(2018)\citenamefont {Qi},
  \citenamefont {Cao}, \citenamefont {Chen}, \citenamefont {Fang},
  \citenamefont {Pan}, \citenamefont {Hao}, \citenamefont {Yao}, \citenamefont
  {Yu},\ and\ \citenamefont {Liu}}]{qi2018effects}%
  \BibitemOpen
  \bibfield  {author} {\bibinfo {author} {\bibfnamefont {J.}~\bibnamefont
  {Qi}}, \bibinfo {author} {\bibfnamefont {M.}~\bibnamefont {Cao}}, \bibinfo
  {author} {\bibfnamefont {Y.}~\bibnamefont {Chen}}, \bibinfo {author}
  {\bibfnamefont {Y.}~\bibnamefont {Fang}}, \bibinfo {author} {\bibfnamefont
  {W.}~\bibnamefont {Pan}}, \bibinfo {author} {\bibfnamefont {H.}~\bibnamefont
  {Hao}}, \bibinfo {author} {\bibfnamefont {Z.}~\bibnamefont {Yao}}, \bibinfo
  {author} {\bibfnamefont {Z.}~\bibnamefont {Yu}}, \ and\ \bibinfo {author}
  {\bibfnamefont {H.}~\bibnamefont {Liu}},\ }\href@noop {} {\bibfield
  {journal} {\bibinfo  {journal} {Journal of Alloys and Compounds}\ }\textbf
  {\bibinfo {volume} {762}},\ \bibinfo {pages} {950} (\bibinfo {year}
  {2018})}\BibitemShut {NoStop}%
\bibitem [{\citenamefont {Fu}\ \emph {et~al.}(2008)\citenamefont {Fu},
  \citenamefont {Liu},\ and\ \citenamefont {Chen}}]{fu2008structure}%
  \BibitemOpen
  \bibfield  {author} {\bibinfo {author} {\bibfnamefont {M.~S.}\ \bibnamefont
  {Fu}}, \bibinfo {author} {\bibfnamefont {X.~Q.}\ \bibnamefont {Liu}}, \ and\
  \bibinfo {author} {\bibfnamefont {X.~M.}\ \bibnamefont {Chen}},\ }\href@noop
  {} {\bibfield  {journal} {\bibinfo  {journal} {Journal of the European
  Ceramic Society}\ }\textbf {\bibinfo {volume} {28}},\ \bibinfo {pages} {585}
  (\bibinfo {year} {2008})}\BibitemShut {NoStop}%
\bibitem [{\citenamefont {Kipkoech}\ \emph {et~al.}(2005)\citenamefont
  {Kipkoech}, \citenamefont {Azough},\ and\ \citenamefont
  {Freer}}]{kipkoech2005microstructural}%
  \BibitemOpen
  \bibfield  {author} {\bibinfo {author} {\bibfnamefont {E.~R.}\ \bibnamefont
  {Kipkoech}}, \bibinfo {author} {\bibfnamefont {F.}~\bibnamefont {Azough}}, \
  and\ \bibinfo {author} {\bibfnamefont {R.}~\bibnamefont {Freer}},\
  }\href@noop {} {\bibfield  {journal} {\bibinfo  {journal} {Journal of Applied
  Physics}\ }\textbf {\bibinfo {volume} {97}},\ \bibinfo {pages} {064103}
  (\bibinfo {year} {2005})}\BibitemShut {NoStop}%
\bibitem [{\citenamefont {Dutta}\ \emph {et~al.}(2007)\citenamefont {Dutta},
  \citenamefont {Sinha},\ and\ \citenamefont
  {Shannigrahi}}]{dutta2007dielectric}%
  \BibitemOpen
  \bibfield  {author} {\bibinfo {author} {\bibfnamefont {A.}~\bibnamefont
  {Dutta}}, \bibinfo {author} {\bibfnamefont {T.~P.}\ \bibnamefont {Sinha}}, \
  and\ \bibinfo {author} {\bibfnamefont {S.}~\bibnamefont {Shannigrahi}},\
  }\href@noop {} {\bibfield  {journal} {\bibinfo  {journal} {Physical Review
  B}\ }\textbf {\bibinfo {volume} {76}},\ \bibinfo {pages} {155113} (\bibinfo
  {year} {2007})}\BibitemShut {NoStop}%
\bibitem [{\citenamefont {Trabelsi}\ \emph {et~al.}(2017)\citenamefont
  {Trabelsi}, \citenamefont {Bejar}, \citenamefont {Dhahri}, \citenamefont
  {Sajieddine}, \citenamefont {Khirouni}, \citenamefont {Prezas}, \citenamefont
  {Melo}, \citenamefont {Valente},\ and\ \citenamefont
  {Gra{\c{c}}a}}]{trabelsi2017effect}%
  \BibitemOpen
  \bibfield  {author} {\bibinfo {author} {\bibfnamefont {H.}~\bibnamefont
  {Trabelsi}}, \bibinfo {author} {\bibfnamefont {M.}~\bibnamefont {Bejar}},
  \bibinfo {author} {\bibfnamefont {E.}~\bibnamefont {Dhahri}}, \bibinfo
  {author} {\bibfnamefont {M.}~\bibnamefont {Sajieddine}}, \bibinfo {author}
  {\bibfnamefont {K.}~\bibnamefont {Khirouni}}, \bibinfo {author}
  {\bibfnamefont {P.~R.}\ \bibnamefont {Prezas}}, \bibinfo {author}
  {\bibfnamefont {B.~M.~G.}\ \bibnamefont {Melo}}, \bibinfo {author}
  {\bibfnamefont {M.~A.}\ \bibnamefont {Valente}}, \ and\ \bibinfo {author}
  {\bibfnamefont {M.~P.~F.}\ \bibnamefont {Gra{\c{c}}a}},\ }\href@noop {}
  {\bibfield  {journal} {\bibinfo  {journal} {Journal of Alloys and Compounds}\
  }\textbf {\bibinfo {volume} {723}},\ \bibinfo {pages} {894} (\bibinfo {year}
  {2017})}\BibitemShut {NoStop}%
\bibitem [{\citenamefont {Smari}\ \emph {et~al.}(2014)\citenamefont {Smari},
  \citenamefont {Rahmouni}, \citenamefont {Elghoul}, \citenamefont {Walha},
  \citenamefont {Dhahri},\ and\ \citenamefont {Khirouni}}]{smari2014electric}%
  \BibitemOpen
  \bibfield  {author} {\bibinfo {author} {\bibfnamefont {M.}~\bibnamefont
  {Smari}}, \bibinfo {author} {\bibfnamefont {H.}~\bibnamefont {Rahmouni}},
  \bibinfo {author} {\bibfnamefont {N.}~\bibnamefont {Elghoul}}, \bibinfo
  {author} {\bibfnamefont {I.}~\bibnamefont {Walha}}, \bibinfo {author}
  {\bibfnamefont {E.}~\bibnamefont {Dhahri}}, \ and\ \bibinfo {author}
  {\bibfnamefont {K.}~\bibnamefont {Khirouni}},\ }\href@noop {} {\bibfield
  {journal} {\bibinfo  {journal} {RSC Advances}\ }\textbf {\bibinfo {volume}
  {5}},\ \bibinfo {pages} {2177} (\bibinfo {year} {2014})}\BibitemShut
  {NoStop}%
\bibitem [{\citenamefont {Jurado}\ \emph {et~al.}(2000)\citenamefont {Jurado},
  \citenamefont {Colomer},\ and\ \citenamefont {Frade}}]{jurado2000electrical}%
  \BibitemOpen
  \bibfield  {author} {\bibinfo {author} {\bibfnamefont {J.~R.}\ \bibnamefont
  {Jurado}}, \bibinfo {author} {\bibfnamefont {M.~T.}\ \bibnamefont {Colomer}},
  \ and\ \bibinfo {author} {\bibfnamefont {J.~R.}\ \bibnamefont {Frade}},\
  }\href@noop {} {\bibfield  {journal} {\bibinfo  {journal} {Journal of the
  American Ceramic Society}\ }\textbf {\bibinfo {volume} {83}},\ \bibinfo
  {pages} {2715} (\bibinfo {year} {2000})}\BibitemShut {NoStop}%
\bibitem [{\citenamefont {Vollman}\ and\ \citenamefont
  {Waser}(1994)}]{vollman1994grain}%
  \BibitemOpen
  \bibfield  {author} {\bibinfo {author} {\bibfnamefont {M.}~\bibnamefont
  {Vollman}}\ and\ \bibinfo {author} {\bibfnamefont {R.}~\bibnamefont
  {Waser}},\ }\href@noop {} {\bibfield  {journal} {\bibinfo  {journal} {Journal
  of the American Ceramic Society}\ }\textbf {\bibinfo {volume} {77}},\
  \bibinfo {pages} {235} (\bibinfo {year} {1994})}\BibitemShut {NoStop}%
\bibitem [{\citenamefont {Vollmann}\ \emph {et~al.}(1997)\citenamefont
  {Vollmann}, \citenamefont {Hagenbeck},\ and\ \citenamefont
  {Waser}}]{vollmann1997grain}%
  \BibitemOpen
  \bibfield  {author} {\bibinfo {author} {\bibfnamefont {M.}~\bibnamefont
  {Vollmann}}, \bibinfo {author} {\bibfnamefont {R.}~\bibnamefont {Hagenbeck}},
  \ and\ \bibinfo {author} {\bibfnamefont {R.}~\bibnamefont {Waser}},\
  }\href@noop {} {\bibfield  {journal} {\bibinfo  {journal} {Journal of the
  American Ceramic Society}\ }\textbf {\bibinfo {volume} {80}},\ \bibinfo
  {pages} {2301} (\bibinfo {year} {1997})}\BibitemShut {NoStop}%
\bibitem [{\citenamefont {Shibata}\ \emph {et~al.}(2007)\citenamefont
  {Shibata}, \citenamefont {Chattopadhyay}, \citenamefont {Lin},\ and\
  \citenamefont {Palkar}}]{shibata2007xafs}%
  \BibitemOpen
  \bibfield  {author} {\bibinfo {author} {\bibfnamefont {T.}~\bibnamefont
  {Shibata}}, \bibinfo {author} {\bibfnamefont {S.}~\bibnamefont
  {Chattopadhyay}}, \bibinfo {author} {\bibfnamefont {B.}~\bibnamefont {Lin}},
  \ and\ \bibinfo {author} {\bibfnamefont {V.~R.}\ \bibnamefont {Palkar}},\
  }in\ \href@noop {} {\emph {\bibinfo {booktitle} {AIP Conference
  Proceedings}}},\ Vol.\ \bibinfo {volume} {882}\ (\bibinfo {organization}
  {American Institute of Physics},\ \bibinfo {year} {2007})\ pp.\ \bibinfo
  {pages} {780--782}\BibitemShut {NoStop}%
\bibitem [{\citenamefont {Chowdhury}\ \emph {et~al.}(2017)\citenamefont
  {Chowdhury}, \citenamefont {Kamal}, \citenamefont {Hossain}, \citenamefont
  {Hasan}, \citenamefont {Islam}, \citenamefont {Ahmmad},\ and\ \citenamefont
  {Basith}}]{chowdhury2017dy}%
  \BibitemOpen
  \bibfield  {author} {\bibinfo {author} {\bibfnamefont {S.~S.}\ \bibnamefont
  {Chowdhury}}, \bibinfo {author} {\bibfnamefont {A.~H.~M.}\ \bibnamefont
  {Kamal}}, \bibinfo {author} {\bibfnamefont {R.}~\bibnamefont {Hossain}},
  \bibinfo {author} {\bibfnamefont {M.}~\bibnamefont {Hasan}}, \bibinfo
  {author} {\bibfnamefont {M.~F.}\ \bibnamefont {Islam}}, \bibinfo {author}
  {\bibfnamefont {B.}~\bibnamefont {Ahmmad}}, \ and\ \bibinfo {author}
  {\bibfnamefont {M.~A.}\ \bibnamefont {Basith}},\ }\href@noop {} {\bibfield
  {journal} {\bibinfo  {journal} {Ceramics International}\ }\textbf {\bibinfo
  {volume} {43}},\ \bibinfo {pages} {9191} (\bibinfo {year}
  {2017})}\BibitemShut {NoStop}%
\bibitem [{\citenamefont {Basith}\ \emph {et~al.}(2015)\citenamefont {Basith},
  \citenamefont {Khan}, \citenamefont {Ahmmad}, \citenamefont {Kubota},
  \citenamefont {Hirose}, \citenamefont {Ngo}, \citenamefont {Tran},\ and\
  \citenamefont {M{\o}lhave}}]{basith2015tunable}%
  \BibitemOpen
  \bibfield  {author} {\bibinfo {author} {\bibfnamefont {M.~A.}\ \bibnamefont
  {Basith}}, \bibinfo {author} {\bibfnamefont {F.~A.}\ \bibnamefont {Khan}},
  \bibinfo {author} {\bibfnamefont {B.}~\bibnamefont {Ahmmad}}, \bibinfo
  {author} {\bibfnamefont {S.}~\bibnamefont {Kubota}}, \bibinfo {author}
  {\bibfnamefont {F.}~\bibnamefont {Hirose}}, \bibinfo {author} {\bibfnamefont
  {D.-T.}\ \bibnamefont {Ngo}}, \bibinfo {author} {\bibfnamefont {Q.-H.}\
  \bibnamefont {Tran}}, \ and\ \bibinfo {author} {\bibfnamefont
  {K.}~\bibnamefont {M{\o}lhave}},\ }\href@noop {} {\bibfield  {journal}
  {\bibinfo  {journal} {Journal of Applied Physics}\ }\textbf {\bibinfo
  {volume} {118}},\ \bibinfo {pages} {023901} (\bibinfo {year}
  {2015})}\BibitemShut {NoStop}%
\bibitem [{\citenamefont {Xu}\ \emph {et~al.}(2013)\citenamefont {Xu},
  \citenamefont {Yang}, \citenamefont {Bai}, \citenamefont {Tang},
  \citenamefont {Zhang},\ and\ \citenamefont {Tang}}]{xu2013oxygen}%
  \BibitemOpen
  \bibfield  {author} {\bibinfo {author} {\bibfnamefont {W.}~\bibnamefont
  {Xu}}, \bibinfo {author} {\bibfnamefont {J.}~\bibnamefont {Yang}}, \bibinfo
  {author} {\bibfnamefont {W.}~\bibnamefont {Bai}}, \bibinfo {author}
  {\bibfnamefont {K.}~\bibnamefont {Tang}}, \bibinfo {author} {\bibfnamefont
  {Y.}~\bibnamefont {Zhang}}, \ and\ \bibinfo {author} {\bibfnamefont
  {X.}~\bibnamefont {Tang}},\ }\href@noop {} {\bibfield  {journal} {\bibinfo
  {journal} {Journal of Applied Physics}\ }\textbf {\bibinfo {volume} {114}},\
  \bibinfo {pages} {154106} (\bibinfo {year} {2013})}\BibitemShut {NoStop}%
\bibitem [{\citenamefont {Verma}\ \emph {et~al.}(2008)\citenamefont {Verma},
  \citenamefont {Kotnala}, \citenamefont {Thakur}, \citenamefont {Rangra},\
  and\ \citenamefont {Negi}}]{verma2008resistivity}%
  \BibitemOpen
  \bibfield  {author} {\bibinfo {author} {\bibfnamefont {K.~C.}\ \bibnamefont
  {Verma}}, \bibinfo {author} {\bibfnamefont {R.~K.}\ \bibnamefont {Kotnala}},
  \bibinfo {author} {\bibfnamefont {N.}~\bibnamefont {Thakur}}, \bibinfo
  {author} {\bibfnamefont {V.~S.}\ \bibnamefont {Rangra}}, \ and\ \bibinfo
  {author} {\bibfnamefont {N.~S.}\ \bibnamefont {Negi}},\ }\href@noop {}
  {\bibfield  {journal} {\bibinfo  {journal} {Journal of Applied Physics}\
  }\textbf {\bibinfo {volume} {104}},\ \bibinfo {pages} {093908} (\bibinfo
  {year} {2008})}\BibitemShut {NoStop}%
\bibitem [{\citenamefont {Zhang}\ \emph {et~al.}(2011)\citenamefont {Zhang},
  \citenamefont {Hu}, \citenamefont {Xu}, \citenamefont {Qin}, \citenamefont
  {Sun}, \citenamefont {Gao}, \citenamefont {Zhang},\ and\ \citenamefont
  {Jiang}}]{zhang2011room}%
  \BibitemOpen
  \bibfield  {author} {\bibinfo {author} {\bibfnamefont {Z.}~\bibnamefont
  {Zhang}}, \bibinfo {author} {\bibfnamefont {J.}~\bibnamefont {Hu}}, \bibinfo
  {author} {\bibfnamefont {Z.}~\bibnamefont {Xu}}, \bibinfo {author}
  {\bibfnamefont {H.}~\bibnamefont {Qin}}, \bibinfo {author} {\bibfnamefont
  {L.}~\bibnamefont {Sun}}, \bibinfo {author} {\bibfnamefont {F.}~\bibnamefont
  {Gao}}, \bibinfo {author} {\bibfnamefont {Y.}~\bibnamefont {Zhang}}, \ and\
  \bibinfo {author} {\bibfnamefont {M.}~\bibnamefont {Jiang}},\ }\href@noop {}
  {\bibfield  {journal} {\bibinfo  {journal} {Solid State Sciences}\ }\textbf
  {\bibinfo {volume} {13}},\ \bibinfo {pages} {1391} (\bibinfo {year}
  {2011})}\BibitemShut {NoStop}%
\bibitem [{\citenamefont {Crandles}\ \emph {et~al.}(2010)\citenamefont
  {Crandles}, \citenamefont {DesRoches},\ and\ \citenamefont
  {Razavi}}]{crandles2010search}%
  \BibitemOpen
  \bibfield  {author} {\bibinfo {author} {\bibfnamefont {D.~A.}\ \bibnamefont
  {Crandles}}, \bibinfo {author} {\bibfnamefont {B.}~\bibnamefont {DesRoches}},
  \ and\ \bibinfo {author} {\bibfnamefont {F.~S.}\ \bibnamefont {Razavi}},\
  }\href@noop {} {\bibfield  {journal} {\bibinfo  {journal} {Journal of Applied
  Physics}\ }\textbf {\bibinfo {volume} {108}},\ \bibinfo {pages} {053908}
  (\bibinfo {year} {2010})}\BibitemShut {NoStop}%
\bibitem [{\citenamefont {Potzger}\ \emph {et~al.}(2011)\citenamefont
  {Potzger}, \citenamefont {Osten}, \citenamefont {Levin}, \citenamefont
  {Shalimov}, \citenamefont {Talut}, \citenamefont {Reuther}, \citenamefont
  {Arpaci}, \citenamefont {B{\"u}rger}, \citenamefont {Schmidt}, \citenamefont
  {Nestler} \emph {et~al.}}]{potzger2011defect}%
  \BibitemOpen
  \bibfield  {author} {\bibinfo {author} {\bibfnamefont {K.}~\bibnamefont
  {Potzger}}, \bibinfo {author} {\bibfnamefont {J.}~\bibnamefont {Osten}},
  \bibinfo {author} {\bibfnamefont {A.~A.}\ \bibnamefont {Levin}}, \bibinfo
  {author} {\bibfnamefont {A.}~\bibnamefont {Shalimov}}, \bibinfo {author}
  {\bibfnamefont {G.}~\bibnamefont {Talut}}, \bibinfo {author} {\bibfnamefont
  {H.}~\bibnamefont {Reuther}}, \bibinfo {author} {\bibfnamefont
  {S.}~\bibnamefont {Arpaci}}, \bibinfo {author} {\bibfnamefont
  {D.}~\bibnamefont {B{\"u}rger}}, \bibinfo {author} {\bibfnamefont
  {H.}~\bibnamefont {Schmidt}}, \bibinfo {author} {\bibfnamefont
  {T.}~\bibnamefont {Nestler}},  \emph {et~al.},\ }\href@noop {} {\bibfield
  {journal} {\bibinfo  {journal} {Journal of Magnetism and Magnetic materials}\
  }\textbf {\bibinfo {volume} {323}},\ \bibinfo {pages} {1551} (\bibinfo {year}
  {2011})}\BibitemShut {NoStop}%
\bibitem [{\citenamefont {Middey}\ \emph {et~al.}(2012)\citenamefont {Middey},
  \citenamefont {Meneghini},\ and\ \citenamefont {Ray}}]{middey2012evidence}%
  \BibitemOpen
  \bibfield  {author} {\bibinfo {author} {\bibfnamefont {S.}~\bibnamefont
  {Middey}}, \bibinfo {author} {\bibfnamefont {C.}~\bibnamefont {Meneghini}}, \
  and\ \bibinfo {author} {\bibfnamefont {S.}~\bibnamefont {Ray}},\ }\href@noop
  {} {\bibfield  {journal} {\bibinfo  {journal} {Applied Physics Letters}\
  }\textbf {\bibinfo {volume} {101}},\ \bibinfo {pages} {042406} (\bibinfo
  {year} {2012})}\BibitemShut {NoStop}%
\bibitem [{\citenamefont {Blundell}(2001)}]{blundell2003magnetism}%
  \BibitemOpen
  \bibfield  {author} {\bibinfo {author} {\bibfnamefont {S.~H.}\ \bibnamefont
  {Blundell}},\ }\href@noop {} {\emph {\bibinfo {title} {Magnetism in Condensed
  Matter}}},\ \bibinfo {edition} {1st}\ ed.\ (\bibinfo  {publisher} {Oxford.
  Univ. Press Inc.},\ \bibinfo {year} {2001})\BibitemShut {NoStop}%
\bibitem [{\citenamefont {Ren}\ \emph {et~al.}(2007)\citenamefont {Ren},
  \citenamefont {Xu}, \citenamefont {Wei}, \citenamefont {Liu}, \citenamefont
  {Hou}, \citenamefont {Du}, \citenamefont {Weng}, \citenamefont {Shen},\ and\
  \citenamefont {Han}}]{ren2007room}%
  \BibitemOpen
  \bibfield  {author} {\bibinfo {author} {\bibfnamefont {Z.}~\bibnamefont
  {Ren}}, \bibinfo {author} {\bibfnamefont {G.}~\bibnamefont {Xu}}, \bibinfo
  {author} {\bibfnamefont {X.}~\bibnamefont {Wei}}, \bibinfo {author}
  {\bibfnamefont {Y.}~\bibnamefont {Liu}}, \bibinfo {author} {\bibfnamefont
  {X.}~\bibnamefont {Hou}}, \bibinfo {author} {\bibfnamefont {P.}~\bibnamefont
  {Du}}, \bibinfo {author} {\bibfnamefont {W.}~\bibnamefont {Weng}}, \bibinfo
  {author} {\bibfnamefont {G.}~\bibnamefont {Shen}}, \ and\ \bibinfo {author}
  {\bibfnamefont {G.}~\bibnamefont {Han}},\ }\href@noop {} {\bibfield
  {journal} {\bibinfo  {journal} {Applied Physics Letters}\ }\textbf {\bibinfo
  {volume} {91}},\ \bibinfo {pages} {063106} (\bibinfo {year}
  {2007})}\BibitemShut {NoStop}%
\bibitem [{\citenamefont {Lei}\ \emph {et~al.}(2014)\citenamefont {Lei},
  \citenamefont {Liu}, \citenamefont {Wang}, \citenamefont {Shen},
  \citenamefont {Guo}, \citenamefont {Wang}, \citenamefont {Zeng},
  \citenamefont {Cheng}, \citenamefont {Xiao},\ and\ \citenamefont
  {Zhou}}]{lei2014ferromagnetic}%
  \BibitemOpen
  \bibfield  {author} {\bibinfo {author} {\bibfnamefont {S.}~\bibnamefont
  {Lei}}, \bibinfo {author} {\bibfnamefont {L.}~\bibnamefont {Liu}}, \bibinfo
  {author} {\bibfnamefont {C.}~\bibnamefont {Wang}}, \bibinfo {author}
  {\bibfnamefont {X.}~\bibnamefont {Shen}}, \bibinfo {author} {\bibfnamefont
  {D.}~\bibnamefont {Guo}}, \bibinfo {author} {\bibfnamefont {C.}~\bibnamefont
  {Wang}}, \bibinfo {author} {\bibfnamefont {S.}~\bibnamefont {Zeng}}, \bibinfo
  {author} {\bibfnamefont {B.}~\bibnamefont {Cheng}}, \bibinfo {author}
  {\bibfnamefont {Y.}~\bibnamefont {Xiao}}, \ and\ \bibinfo {author}
  {\bibfnamefont {L.}~\bibnamefont {Zhou}},\ }\href@noop {} {\bibfield
  {journal} {\bibinfo  {journal} {CrystEngComm}\ }\textbf {\bibinfo {volume}
  {16}},\ \bibinfo {pages} {1322} (\bibinfo {year} {2014})}\BibitemShut
  {NoStop}%
\bibitem [{\citenamefont {Ahmmad}\ \emph
  {et~al.}(2016{\natexlab{b}})\citenamefont {Ahmmad}, \citenamefont {Kanomata},
  \citenamefont {Koike}, \citenamefont {Kubota}, \citenamefont {Kato},
  \citenamefont {Hirose}, \citenamefont {Billah}, \citenamefont {Jalil},\ and\
  \citenamefont {Basith}}]{ahmmad2016large}%
  \BibitemOpen
  \bibfield  {author} {\bibinfo {author} {\bibfnamefont {B.}~\bibnamefont
  {Ahmmad}}, \bibinfo {author} {\bibfnamefont {K.}~\bibnamefont {Kanomata}},
  \bibinfo {author} {\bibfnamefont {K.}~\bibnamefont {Koike}}, \bibinfo
  {author} {\bibfnamefont {S.}~\bibnamefont {Kubota}}, \bibinfo {author}
  {\bibfnamefont {H.}~\bibnamefont {Kato}}, \bibinfo {author} {\bibfnamefont
  {F.}~\bibnamefont {Hirose}}, \bibinfo {author} {\bibfnamefont
  {A.}~\bibnamefont {Billah}}, \bibinfo {author} {\bibfnamefont {M.~A.}\
  \bibnamefont {Jalil}}, \ and\ \bibinfo {author} {\bibfnamefont {M.~A.}\
  \bibnamefont {Basith}},\ }\href@noop {} {\bibfield  {journal} {\bibinfo
  {journal} {Journal of Physics D: Applied Physics}\ }\textbf {\bibinfo
  {volume} {49}},\ \bibinfo {pages} {265003} (\bibinfo {year}
  {2016}{\natexlab{b}})}\BibitemShut {NoStop}%
\end{thebibliography}%
\newpage
\clearpage
 
\pagenumbering{arabic}
\thispagestyle{empty}
\begin{figure*}
\begin{center}
    \includegraphics[page=1, scale=.9]{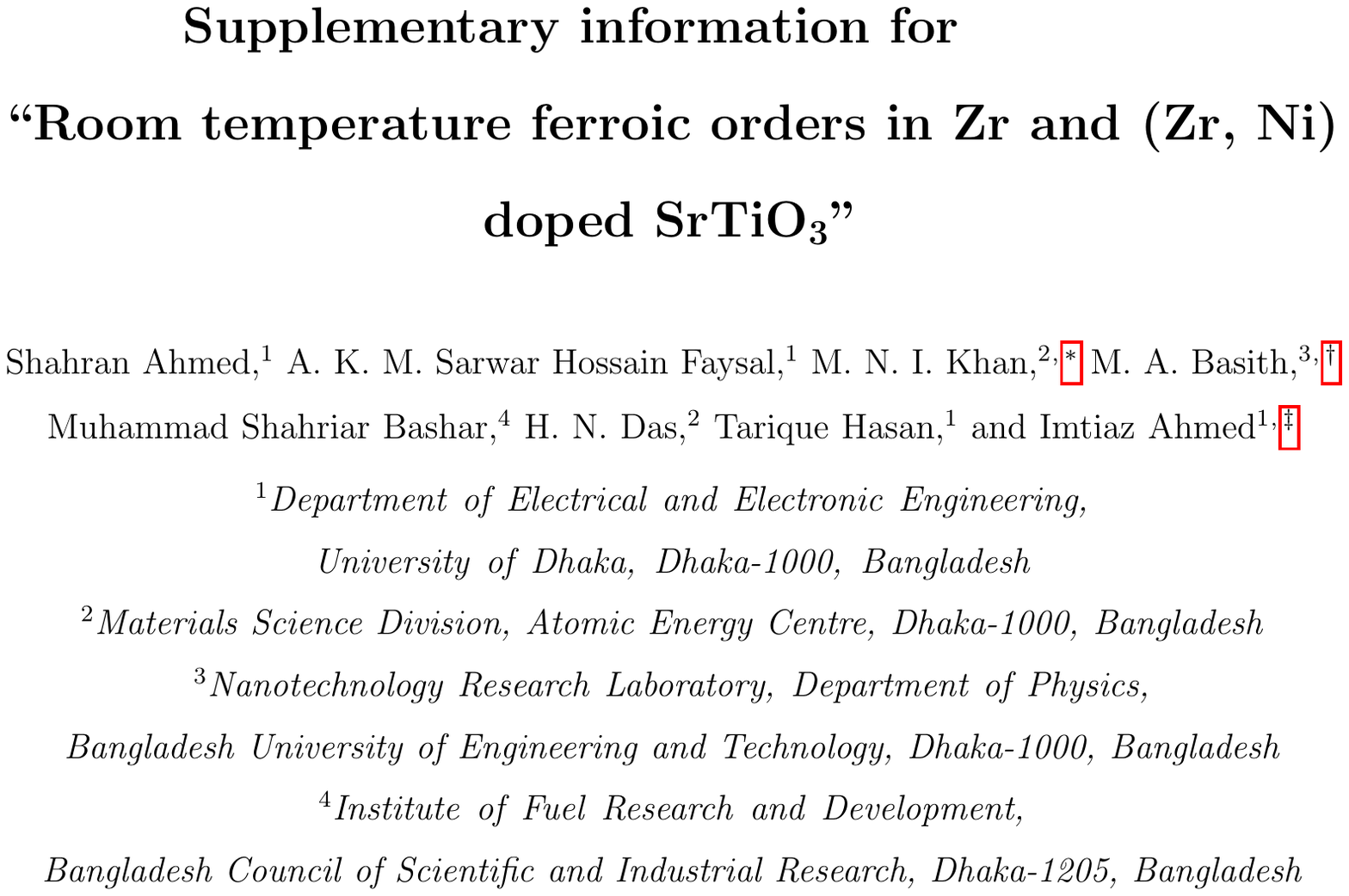}
    \end{center}
\end{figure*}
\begin{figure*}
\begin{center}
    \includegraphics[page=2, scale=0.9]{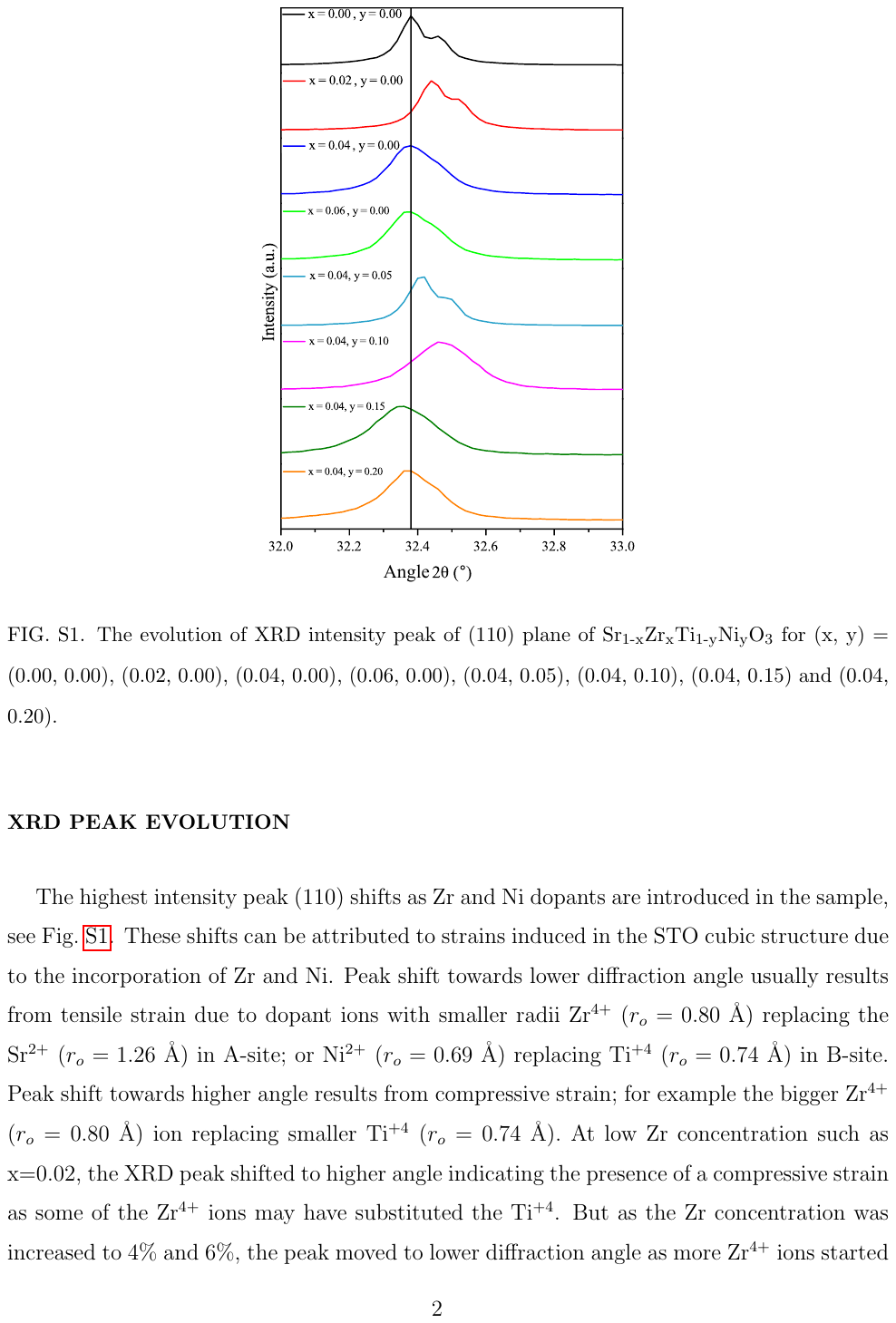}
    \end{center}
\end{figure*}
\begin{figure*}
\begin{center}
    \includegraphics[page=3, scale=0.9]{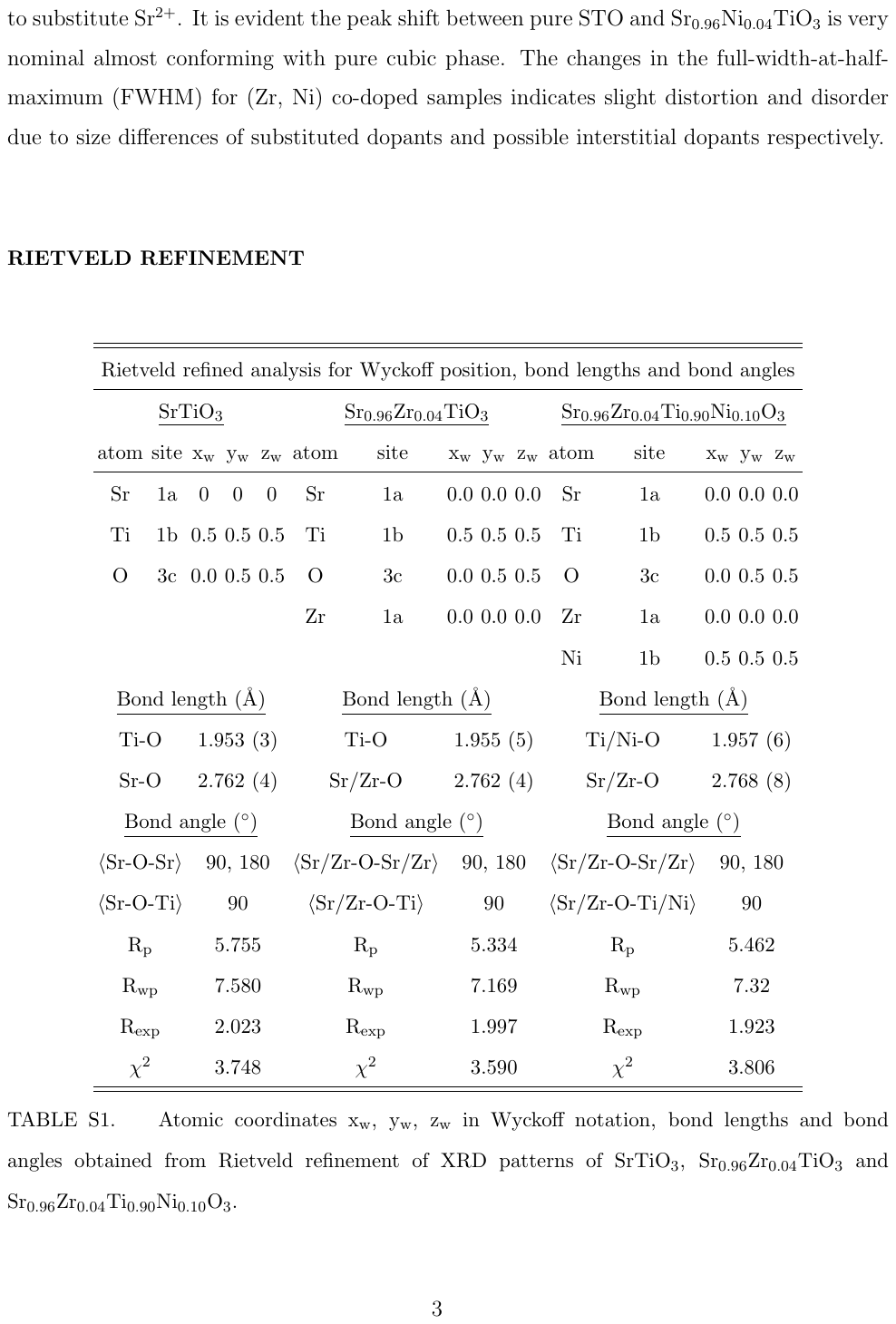}
    \end{center}
\end{figure*}
\begin{figure*}
\begin{center}
    \includegraphics[page=4, scale=0.9]{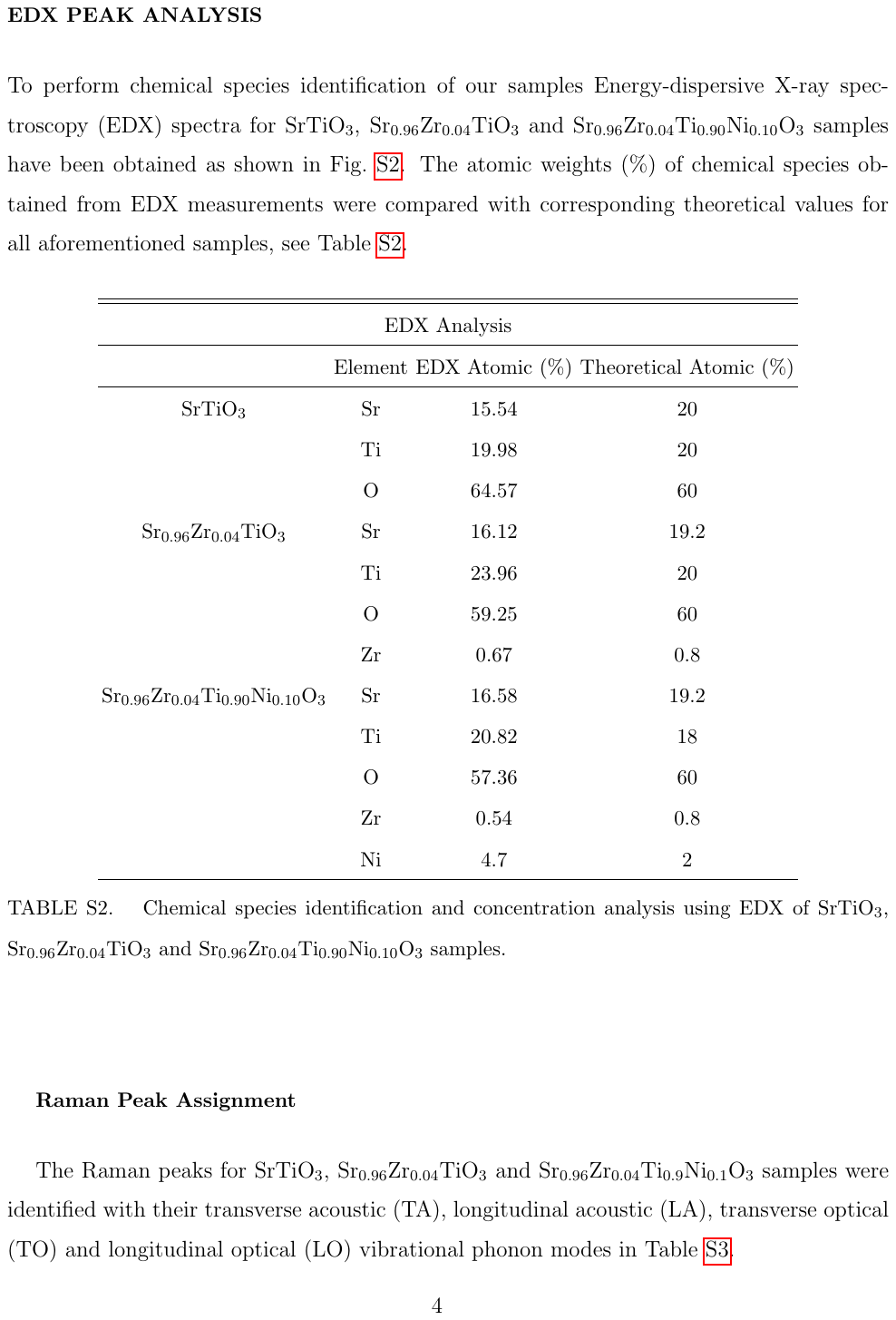}
    \end{center}
\end{figure*}
\begin{figure*}
\begin{center}
    \includegraphics[page=5, scale=0.9]{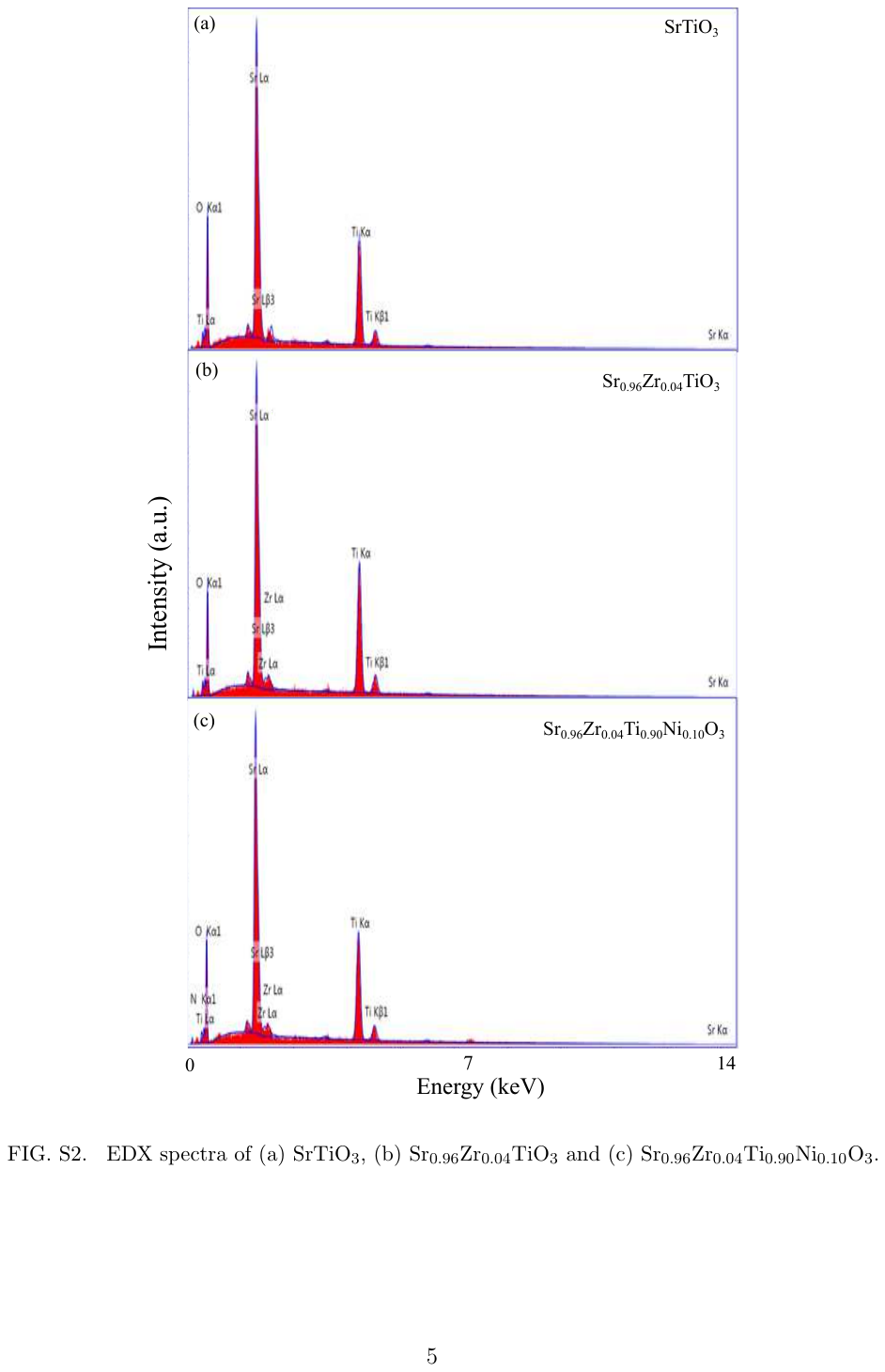}
    \end{center}
\end{figure*}
\begin{figure*}
\begin{center}
    \includegraphics[page=6, scale=0.9]{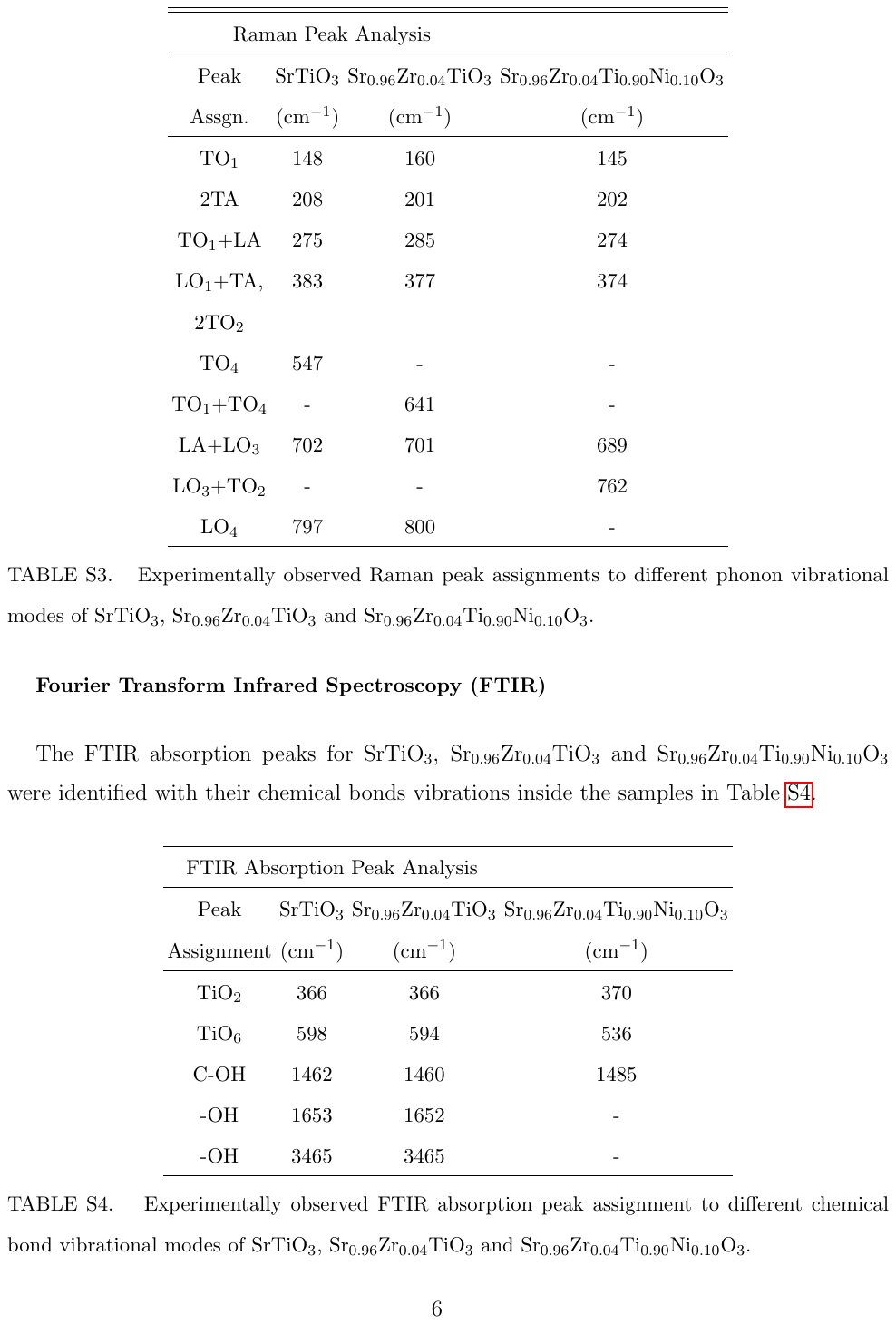}
    \end{center}
\end{figure*}
\begin{figure*}
\begin{center}
    \includegraphics[page=7, scale=0.9]{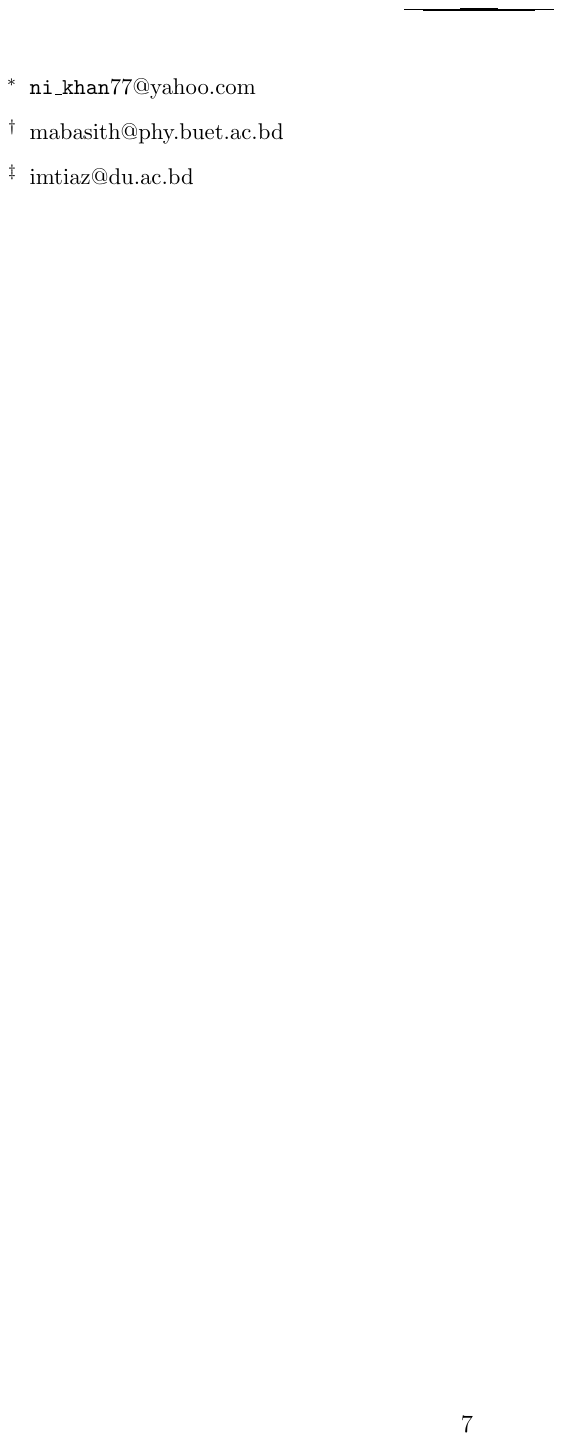}
    \end{center}
\end{figure*}

\end{document}